\documentclass[aps,prb]{revtex4}
\usepackage{bm}
\usepackage{amsmath}
\usepackage{color}
\usepackage{graphicx}
\usepackage{amssymb}
\DeclareMathAlphabet{\mathbbmsl}{U}{bbm}{m}{sl}
\makeatletter
\newsavebox{\@brx}
\newcommand{\llangle}[1][]{\savebox{\@brx}{\(\m@th{#1\langle}\)}%
	\mathopen{\copy\@brx\kern-0.5\wd\@brx\usebox{\@brx}}}
\newcommand{\rrangle}[1][]{\savebox{\@brx}{\(\m@th{#1\rangle}\)}%
	\mathclose{\copy\@brx\kern-0.5\wd\@brx\usebox{\@brx}}}
\makeatother

\begin{document}
\draft
\title{Influence of Dynamical Floquet Spectrum on the Plasmon Excitations  and Exchange Energy of  tilted monolayer  1T$^\prime$MoS$_2$}

\author{Sita Kandel$^{1,2}$,   Godfrey Gumbs$^{1,2,3}$ , Antonios Balassis$^{4}$, Andrii Iurov$^5$ and Oleksiy Roslyak$^{4}$ }
\address{$^1$Department of Physics, Hunter College, City University of New York, 695 Park Avenue, New York, NY 10065 USA}
\address{$^{2}$The Graduate School and University Center, The
City University of New York,    New York, NY 10016, USA}
\address{$^3$Donostia International Physics Center (DIPC), P de Manuel Lardizabal, 4, 20018 San Sebastian, Basque Country, Spain  }
\address{$^{4}$Department of Physics and Engineering Physics, Fordham University, 441 East Fordham Road, Bronx, 10458, NY, USA}
\address{$^{5}$Department of Physics and Computer Science, Medgar Evers College of City University of New York, Brooklyn, NY 11225, USA} 

\date{\today}

\begin{abstract}
It is now well established that a high-frequency electromagnetic dressing field within the off-resonance regime can significantly  modify the electronic, transport and optical properties of Dirac materials. Here,  using light with circular polarization and in the terahertz range, we investigate its effect on the energy spectrum (dressed states), polarization function, plasmons and  their lifetimes due to Landau damping of tilted monolayer  1T$^\prime$MoS$_2$ which acquires two energy gaps associated with up- and down- pseudospin.   We can adjust its electronic properties over a wider range by varying these two band gaps in contrast with graphene. Unlike graphene and other two dimensional materials, the plasmons are Landau damped in the long wavelength limit,  and emerge from  the particle-hole continuum at finite  wave vector $q$ and frequency $\omega$ even in the absence of irradiation. This damping arises from intraband transitions. With the use of the  Lindhard approach for the frequency-dependent polarizability propagator, we have developed a rigorous theoretical formalism for employing the Floquet energy spectrum for investigating the many-body effects on the plasmon excitations,  their lifetimes due to Landau damping  and the exchange energy of tilted monolayer 1T$^\prime$MoS$_2$ under normal incidence of electromagnetic radiation at arbitrary temperature. The dressed states at very low temperature corresponding to circular polarization suppress the response of the system to the external probe.  In the long wavelength limit, the, $\sqrt{qT}$-dependent plasmons is  restored at high temperatures. Our calculations have shown that the exchange energy decreases exponentially as a function of chemical potential for finite doping.  Its behavior is also   influenced  by the indirect momentum space pseudo spin gap, the tilting of the energy bands, and anisotropy.  The results for the exchange could be useful in  calculate the compressibility and scattering rates of  the tilted gapped monolayer Dirac material 1T$^\prime$MoS$_2$.
\end{abstract}

\maketitle

\medskip

\medskip

\noindent
{\bf Corresponding author}:\ \     Sita Kandel; E-mail: skandel@gradcenter.cuny.edu.

\section{Introduction}
\label{sec1}

\medskip 
\par
There has been a tremendous amount of research into the properties of two-dimensional  (2D) materials with a Dirac cone band structure following  the pioneering transport measurements  on graphene.  There is now a copious amount of data from  studies regarding their various electronic and optoelectronic properties.  The relevant materials include recently discovered   $\alpha$-T$_3$ model with a flat band, \cite{AI1,AI2,AI3,AI4,AI5,AI6,AI7}  anisotropic and tilted 1T$^\prime$MoS$_2$,  \cite{AI8,AI9,AI10,AI11}  semi-Dirac materials \cite{AI12,AI13,AI14}   and materials with significant Rashba spin-orbit coupling.  \cite{AI15,AI16}   An  interesting feature of some of these Dirac cone materials is the anisotropy and energy gap in their band  structure. The band gap may be induced  in a variety of ways including by applying external  off-resonance irradiation\cite{AI17,AI18}   which is a subject of interest in this paper.

\begin{figure}[ht]
\centering
\includegraphics[width=0.45\linewidth]{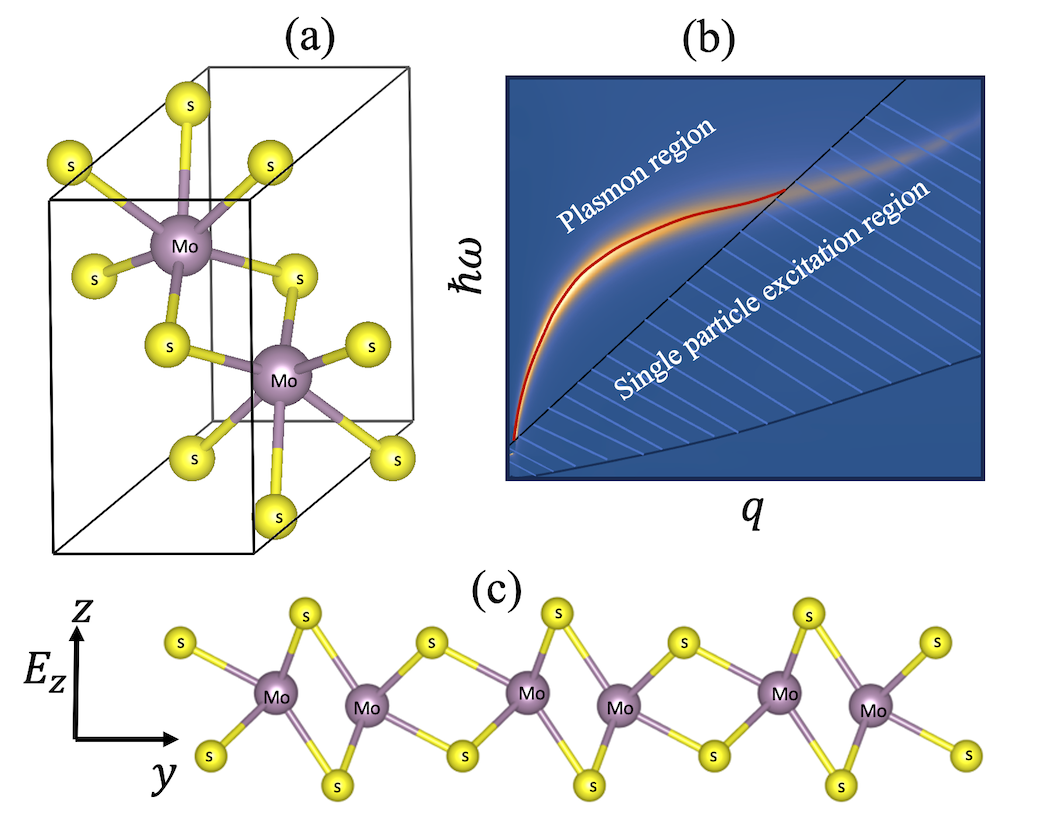}
\caption{ (Color online) Schematic of the crystal structure of 1T$^\prime$MoS$_2$ showing both  the 3D  lattice structure in (a) and the  sideways view  depicting a buckled structure in (c), where an electric field $E_{z}$ is applied along the perpendicular $z$ direction.   The brown disks are for Mo and the yellow disks represent S atoms. Schematic representations of  plasmon dispersion  for  1T$^\prime$MoS$_2$ along one wave vector direction is presented in (b). The shaded region is the single-particle excitation region, where the plasmon is  Landau  damped. This damping occurs in the long wavelength limit, thereby affecting the square root of $q$ behavior of the self-sustained plasma oscillations.} 
\label{schm}
\end{figure}

\medskip
\par
A class of these Dirac cone materials which have been gaining considerable attention due to their remarkable optoelectronic as well as transport properties are  transition metal dichalcogenides (TMDCs). It has been found that  TMDCs have  applications for a range of fundamental  phenomena.  \cite{ANA1}  Interestingly,  it has been demonstrated that  a double layer of these materials can  support indirect momentum space dark excitons with massive anisotropic tilted semi-Dirac bands.  This has been the subject of recent experimental studies. \cite{ANA4,ANA5,ANA6}   As a consequence of  their thermodynamic stability,  1T$^\prime$Mo$_2$     with its significant anisotropic tilted Dirac  energy bands has been   one of the most interesting 2D Dirac materials.   Its appeal is also the ease of their fabrication in the semiconducting phase.  Additionally,  it  was predicted  that  1T$^\prime$MoS$_2$ has a substantial  quantum-spin Hall effect.\cite{ANA7}   We also note that when a uniform perpendicular electric  field is applied, 1T$^\prime$MoS$_2$ presents valley-spin polarized Dirac bands \cite{ANA8}  and also   a  phase transition from a topological  insulator to a regular band insulator. \cite{AI10}  The anisotropy of the energy band structure gives rise to crucial properties which stem from its tilted  and shifted valence and conduction Dirac bands.  \cite{ANA410}  Figure \  \ref{schm} shows a schematic representation of the lattice structure of 1T$^\prime$MoS$_2$ in the presence of a    vertical electric field as well as an example of the plasmon dispersion along a momentum transfer direction measured from a symmetry K point. The plasmon is Landau damped in the long wavelength regime. 

\medskip
\par
The electron band structure plays a key role in determining the electronic,\cite{Ando,Guinea} optical,\cite{IOP}  Boltzmann transport,\cite{Gulley} plasmonic,\cite{ANA8,plas1,plas2,plas3,plas4}  excitonic\cite{Nafis} properties and the Berry phase\cite{Berry}  of condensed matter.  The role played by the many-body  physics for these  properties is a crucial issue which has caught the attention of many researchers over the years.   At the  center of these investigations is  the role played by exchange and correlation effects.\cite{EXC1}  In this paper, we are interested in contrasting one of  these two  fundamental quantities in Dirac cone materials.

\medskip
\par
With this stated background,  we now turn our attention to describing the  motivation for this  work.   It  is to examine the impact of irradiation  on the band structure  (dressed states) anisotropy  and the band gaps on major many-body properties, namely the collective  excitations (plasmons)\cite{silkin}, their lifetimes due to Landau damping, the polarizability and the exchange energy in 1T$^\prime$ TMDC within the framework of the random-phase approximation (RPA). We also examine the role played by temperature  on these crucial many-body properties. The dynamical polarization function is an important  and fundamental quantity which is employed to  help understand these properties.\cite{ SK38, SK19,SK20,SK21,SK22,SK23,SK24,SK25,SK26} On one hand the plasmon modes calculated and analyzed in our study hold promise for successful experimental probing through techniques such as inelastic angle resolved photoemission spectroscopy (ARPES), electron scattering or high resolution electron energy-loss spectroscopy (HREELS) \cite{PRB9,PRB10}.  High resolution EELS has emerged as the conventional approach for investigating collective Coulomb excitations, namely plasmons. Notably, HREELS has been instrumental in exploring plasmon dispersion on epitaxial graphene. \cite{PRB11}  Over time, this methodology has undergone enhancements, achieving subangstrom spatial resolution,  thereby facilitating insights into interband transitions between valence and conduction bands, as well as core-level excitations.  On the other hand,  calculations of quantum  mechanical interaction and many-body screening, pertaining to  the charge and spin susceptibilities, usually require computing the exchange and correlation energies of conduction electrons which may be based on the RPA.  The compressibility due to the exchange contribution was first calculated in Ref.\ [\onlinecite{SK30}] for a graphene flake. This predicted local compressibility of graphene has been experimentally verified\cite{SK31}  using a scanning single electron transistor.  The same effect due to the exchange interaction between spins has also been found to be important when assessing electron entanglement\cite{SK32} in quantum-computing applications.    

\medskip
\par

Recent progress in optical and microwave physics, advancements in laser technology, and the emergence of technical applications in condensed matter quantum optics have facilitated the  practical implementation of Floquet engineering in real-world optoelectronic devices. \cite{PRB 5,PRB6,PRB6a,PRB7} A challenge in the application of Floquet theory is heating which arises in the driven system. Efforts  have been made to address  effects due to heating. These include exploring non-ergodic systems and utilizing destructive interference of excitation pathways through a two-color driving mechanism.  By controlling heating, one could access strongly correlated phases like fractional quantum Hall states and half-integer Mott insulators.  Additionally, it would extend quantum simulation to high-energy concepts such as dynamical gauge fields. \cite{PRB8}  

\medskip
\par

An application of interest for the exchange energy has been to the scattering rates of the Coulomb excitations.\cite{GG2} There, the deexcitation processes were studied using the screened exchange and correlation energy. That calculation  employed the intraband single-particle excitations (SPEs), the interband SPEs, and the plasmon modes, depending on the quasiparticle states and the Fermi energies.

\medskip
\par
The   remainder of this paper is organized in the  following way.  In  Sec.\  \ref{sec2}(A),   we introduce a low-energy Hamiltonian for 1T$^\prime$MoS$_2$ in the absence of irradiation but in the presence of a perpendicular electric field and briefly  review some essential properties resulting  from dispersion and the corresponding electronic states in the absence of irradiation.  Section \ref{sec2}(B)  is devoted to deriving  an effective Hamiltonian when the system is irradiated by circularly polarized light.  We also 
analyze the electron-photon dressed states under this irradiation.    
In Sec. \  \ref{sec3}.  we calculate the intraband and interband contributions to the polarization function and investigate the  dependence of the total polarization function  for chosen doping level  as well as  perpendicular electric field and compare these results in the presence and absence of circularly polarized light.  With the use of the derived polarization function, we  obtain the plasmon dispersion relation as well as its damping rate and demonstrate that these rates are affected by  irradiation when the wavelength of the plasmon mode is much shorter than that of the circular polarized light.  The temperature-induced  plasmon excitation spectra for 1T$^\prime$MoS$_2$ are calculated and analyzed in Sec. \ref{sec4}.   Our results for the exchange energy are given in Sec. \ref{sec5}.  The final conclusions of our presented results and  outlook are presented in Section \ref{sec6}.   

\medskip
\par

\section{Low-energy Model Hamiltonian}
\label{sec2}

\subsection{Electron states in 1T$^\prime$MoS$_2$ }

The first  step  in our calculation is the construction of the low-energy model Hamiltonian for 1T$^\prime$MoS$_2$ monolayer. This is achieved with the use of the   $\mathbf {k}\cdot {\mathbf p}$ method based on the symmetry properties of the valence and conduction bands of the MoS$_2$ crystalline structure. The valence and conduction bands consist primarily of $d$-orbitals of Mo atoms and by $p_y$-orbitals of S atoms, respectively.  The notation we use is as follows:  $\lambda = \pm$ is used to distinguish the locations of two independent Dirac points. $\upsilon_1 = 3.87 \times 10^5$ m/s and $\upsilon_2 = 0.46 \times 10^5$ m/s denote the Fermi velocities along the $x$ and $y$ directions, respectively.  $\upsilon_{-} = 2.86 \times 10^5$ m/s and $\upsilon_2 = 7.21 \times 10^5$ m/s are the velocity correction terms around the two Dirac points. The $4 \times 4$ unit matrix ${\mathbf{ I}} = \tau_0 \otimes \sigma_0$ and Dirac matrices $\gamma_0 = \tau_1 \otimes \sigma_1$,  $\gamma_1 = \tau_2 \otimes \sigma_0$,  $\gamma_2 = \tau_3 \otimes \sigma_0$ are defined based on the pseudospin space  $\tau_{0, 1, 2, 3}$ and Pauli matrices $\sigma_{0, 1, 2, 3}$. In addition, ${\bf k} = (k_x, k_y)$  is the wave vector,  and $\alpha =  |E_z/E_c|$ is the ratio of the vertical electric field over its critical value for which the band gap is closed.  The SOC gap  depends on the doping concentration of 2H-MoS$_2$ with alkali atoms  and has a typical value of  $\Delta \sim  0.042$ eV.

\medskip
\par
Making use of the above described results,   the low-energy Hamiltonian for a 2D anisotropic tilted Dirac system representing 1T$^\prime$ -MoS$_2$ in the vicinity of two independent Dirac points  located at (0, $\lambda$) with $\lambda=\pm1$ is given by

\begin{equation}
\hat{{\cal H}}_\lambda(k_x,k_y) =  
 \left(
\begin{array}{cccc}
- \hbar (\upsilon_+ + \upsilon_-)\,\lambda k_y & 0 & \alpha \Delta  - i \hbar \upsilon_1 k_x & \lambda \Delta + \hbar \upsilon_2 k_y \\
0 & -\hbar  (\upsilon_+ + \upsilon_-)\,\lambda  k_y &  \lambda  \Delta +  \hbar \upsilon_2 k_y & \alpha \Delta - i \hbar \upsilon_1 k_x \\  0 & 0 &  \hbar (\upsilon_+ - \upsilon_-)\,\lambda  k_y & 0 \\
0 & 0 & 0 &  \hbar (\upsilon_+ - \upsilon_-)\, \lambda  k_y
\end{array}
\right) + H.c. \, .
\label{EIV}
\end{equation}
      This is the general form presented in Ref.\ [\onlinecite{SK34}]  for an effective low-energy Hamiltonian  near the Dirac point for  one of the tilted Dirac cones  as   a $2\times 2$ matrix in which the tilt is represented by the velocity component corrections around the two Dirac points.

\medskip
\par
Calculation shows that the energy eigenvalues of  Eq.\ (\ref{EIV}) are given by\cite{AI10}

\begin{equation}
   E_{\xi,s}^{\lambda}({\bf k})= E_{\xi,s}^{\lambda}(k_x,k_y)=-\lambda \hbar \upsilon_{-} k_{y}+\xi \sqrt{\left[\hbar \upsilon_{2} k_{y}+\left(\lambda-s\alpha\right)\Delta  \right]^2+\left(\hbar \upsilon_{1} k_{x}\right)^2+\left(\hbar \upsilon_{+} k_{y}\right)^2} \  ,
\label{Eigen_1}
\end{equation}
where $\xi=\pm 1$ for the conduction (valence) band, $s=\pm 1$ is the spin up (down) index. 
It is clear from Eq.\,(\ref{Eigen_1}) that the energy spectrum becomes anisotropic and tilted with respect to the $k_y$ axis,  i.e., the $\,\lambda \hbar \upsilon_- k_y$ term gives rise to an equal-contribution in magnitude  but with opposite sign around $k_y = 0$.   Additionally, the energy band gap is indirect and closes when  $\lambda- s\,\alpha=0$ is satisfied. 

\medskip
\par

\begin{figure}[!ht]
\centering
\includegraphics[width=0.7\linewidth]{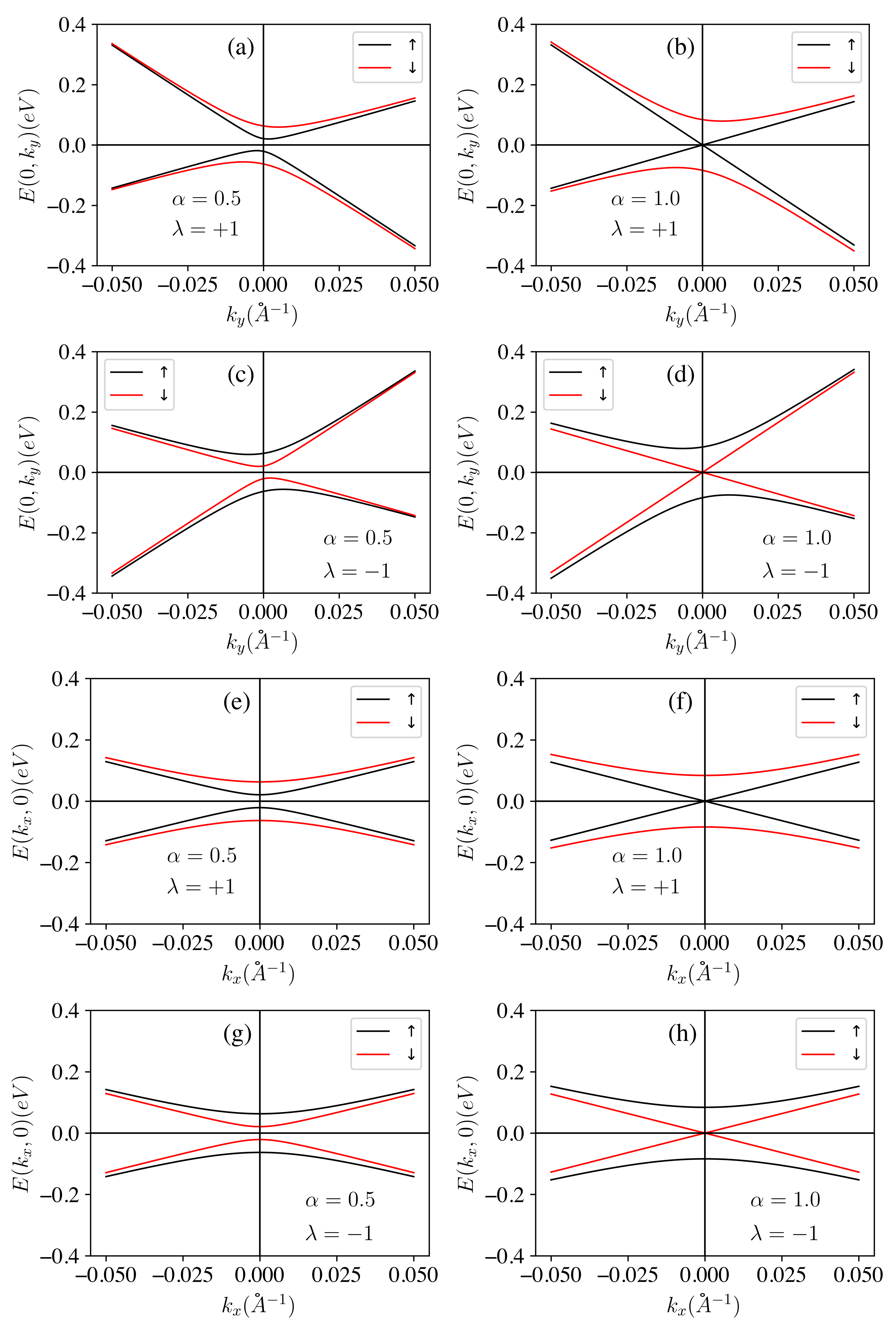}
\caption{ (Color online)  Energy dispersion of tilted 1T$^{\prime}$MoS$_2$ lattice, in the presence of a vertical electric field, as functions of momentum transfer $k_y$ in panels (a),(b),(c) and (d) when $k_x=0$.  The plots  in (e),(f),(g) and (h) are the corresponding results as functions of $k_x$  when $k_y=0$. The panels on the right-hand side for each chosen axis correspond to the application of a critical vertical electric field ($\alpha$ = 1), whereas  the panels on the left for each axis are  for an electric field equal to half the critical value, i.e.,  $\alpha$ = 0.5. The dispersions at  both Dirac points corresponding to $\lambda = \pm1$ are presented as indicated in the figure. }
\label{energy_1}
\end{figure}

\medskip
\par 

Our calculations also show that the wave functions   for the conduction and valence bands ($\xi=\pm 1$) are valley-dependent  ($\lambda=\pm 1$) as well as spin-dependent ($s=\pm 1$) and given by the following  expression

\begin{equation}
\psi_{\xi,s}^{\lambda}  ({\bf k},{\bf r})=\left(
\begin{array}{c}
\dfrac{-\lambda \hbar k_{y} \upsilon_{+}+\xi \sqrt{|{\cal D}|^{2}+\left(\hbar k_{y} \upsilon_{+}\right)^{2}}}{{\cal D}} \\[6pt]
\dfrac{\lambda \hbar k_{y} \upsilon_{+}-\xi \sqrt{|{\cal D}|^{2}+\left(\hbar k_{y} \upsilon_{+}\right)^{2}}}{s{\cal D}}) \\[6pt]
-s\\[6pt]
1
\end{array} \right)
\frac{e^{i{\bf k}\cdot {\bf r}}}{\sqrt{A}}, 
\end{equation}
where $A$ is a normalization area and we have introduced  ${\cal D}\equiv  (\lambda-s \alpha) \Delta+\hbar \upsilon_{2} k_{y}-  i s \hbar \upsilon_{1} k_{x}$. For convenience, each of the Fermi velocities  $\upsilon_1, \upsilon_2,\upsilon_{+}, \upsilon_{-}$ is scaled by the Fermi velocity $\upsilon_{-}$ , the wave vector by $k_y^{(0)}=1.0\AA^{-1}$ and energy by $\hbar\upsilon_{-}k_y^{(0)}=1.88$ \ eV.

\medskip
\par 
\noindent 
The two dimensional 1T$^{\prime}$MoS$_2$ phase exhibits anisotropic tilted energy bands along the $k_{y}$ direction in the presence of a vertical electric field, described by a parameter $\alpha \neq 0$, and the spin-orbit coupling gap $\Delta$. Whereas the bands along the $k_{x}$ direction are not tilted and are symmetric about $k_{x} =$ 0.  When the external electric field  is not equal to its critical value, i.e. when $\alpha \neq 1$, the energy bands exhibit a valley-spin polarized tilted energy gap given by $\Delta_{s}^{\lambda} $ = $|\lambda- \alpha s | \Delta$. However, in the absence of  an external electric field (i.e., $\alpha=0$), the energy bands are spin degenerate. It is interesting to note that spin-up bands and spin-down bands split in one valley but reverse their spins in the other valley.   When the electric field is equal to the critical field (i.e., $\alpha=1$), the spin-up bands become gapless in the $\lambda=+1$ valley.  However, in the other valley $\lambda=-1$, the spin-down bands become gapless. Further increase in the field beyond its critical value, the gapless bands separate and a gap appears. The tilted energy dispersion of 1T$^{\prime}$ MoS$_2$ plotted in Fig.\  \ref{energy_1} illustrates that the gap depends on the chosen vertical electric field. In the panels on the left-hand side, when the field  is equal to half the critical value (i.e., $\alpha=0.5$), the finite $\Delta_{s}^{\lambda} $ for both spin channels while that in the right-hand column for the field equal to the  critical value (i.e., $\alpha=1$), the spin up subbands crosses and the spin down subbands are separated by  $2\Delta$. For the typical value of $\Delta \sim 0.042 eV$, the electric field required to close this gap is approximately about $0.06$ V/nm. The tilted bands in $MoS_2$ have many implications in their electronic and optical properties. For example, tilted bands due to  a strain effect resulted in the anisotropic electron mobility by enhancing mobility along specific directions \,\cite{Kalantar}  and tilted Dirac bands in 1T$^{\prime}$ MoS$_2$  developed the anisotropic longitudinal optical conductivity\, \cite{AI10} as well as the anisotropic polarization and plasmon dispersion\cite{ANA8}. We note that the nature of the plots for graph is similar in both  valleys $\lambda =\pm$.  The only difference is that the spins are reversed and the bands are tilted in opposite momentum direction. For the further calculation we added the contribution from both valley.

\subsection{Electron dressed states}

 \medskip
\par
In this section, our objective is computing and investigating electronic states arising from the interaction between electrons and photons.  These states evolve from  the influence of a high-frequency external optical field. Conventionally, we refer to these altered electronic states as ``electron dressed states", with the optical field called the ``dressing field". Quantum mechanically, the vector potential of this dressing field is incorporated into an effective Hamiltonian via a canonical substitution for the electron wave vector ${\bf k} \Rightarrow {\bf k} - q {\bf A}(t)/\hbar$, where $q$ signifies the particle charge (electron/hole), and the time periodic vector potential field ${\bf A}(t)$ is ascertainable by its polarization and the applied electric field strength. To tackle the corresponding eigenvalue problem, we rely on a perturbation technique based on the Floquet-Magnus expansion for an interaction Hamiltonian in terms of powers of inverse frequency. This approach is particularly effective for off-resonant high-frequency irradiation cases.
\medskip
\par 

In general, $\mbox{\boldmath$A$}^{(E)}(t)$ for circular polarization can be written as

\begin{equation}
\label{ellipA}
\mbox{\boldmath$A$}^{(E)}(t) =
\left[  \begin{array}{c}
          A^{(E)}_x (t) \\
          A^{(E)}_y (t)
        \end{array}
\right] = \frac{E_0}{\Omega} \left[
\begin{array}{c}
 \cos (\Omega t + \Theta_p)  \\
\sin (\Omega t + \Theta_p) 
\end{array}
\right] \, 
\end{equation}
where $\Omega$ is the pump frequency and $E_0$ is the strength of the electric field,   $\Theta_p$ denotes the polarization angle, measured from the $x$ axis, of the optical field. With this circularly polarized dressing field, the interaction Hamiltonian is written as  
 
\begin{eqnarray}
\nonumber
 \hat{\bm{H}}_A^{(C)}(t)
 &=& \,\frac{\upsilon_{F} e E_0} {\Omega} \lambda\, \left[(\upsilon_- \, \mbox{\boldmath$\Gamma$}^{(0,0)} + \upsilon_+ \, \mbox{\boldmath$\Gamma$}^{(3,0)}) \, \sin (\Omega t)  -
\upsilon_2 \, \mbox{\boldmath$\Gamma$}^{(1,1)} \,  \sin (\Omega t)  - \upsilon_1\, \mbox{\boldmath$\Gamma$}^{(2,0)} \,  \cos (\Omega t)
\right] \, 
 \label{HAC}
\end{eqnarray}
in terms of the Fermi velocities $\upsilon_F,\upsilon_\pm,\upsilon_{1,2} $, $4 \times 4$ gamma matrices $\mbox{\boldmath$\Gamma$}^{(0,0)} = \mbox{\boldmath$\tau$}_0 \otimes\mbox{\boldmath$\sigma$}_0$, $\mbox{\boldmath$\Gamma$}^{(1,1)} = \mbox{\boldmath$\tau$}_1 \otimes\mbox{\boldmath$\sigma$}_1$,  $\mbox{\boldmath$\Gamma$}^{(2,0)} = \mbox{\boldmath$\tau$}_2 \otimes\mbox{\boldmath$\sigma$}_0$ and $\mbox{\boldmath$\Gamma$}^{(3,0)} = \mbox{\boldmath$\tau$}_3 \otimes\mbox{\boldmath$\sigma$}_0$, where the symbol $\otimes$ represents an outer  product (or Kronecker product) and $\mbox{\boldmath$\tau$}_i$ and $\mbox{\boldmath$\sigma$}_i$ with $i=0,\,1,\, 2,\, 3$ are regular $2 \times 2$ Pauli matrices defined in pseudospin and  real-spin spaces, correspondingly.  Also,  we introduce the notation  $\zeta \equiv \frac{ \upsilon_{F} e E_0/ \Omega}{\hbar \Omega}$ to represent the light-electron interaction energy. 

 \medskip
\par
In the presence of an external driving field the system can be describe by a time-dependent  Hamiltonian given by
\begin{equation}
 \hat{H}^{T}_\lambda(t)  = \hat{{\cal H}}_\lambda(k_x,k_y) +  \hat{\bm{H}}_A^{(C)}(t) 
 \end{equation} 
 where $ \hat{{\cal H}}_\lambda(k_x,k_y) $ is time-independent unperturbed Hamiltonian and $\hat{\bm{H}}_A^{(C)}(t) $ is time-dependent interaction energy. This Hamiltonian is time periodic with period $T = 2\pi/\Omega$ and follow the time-dependent Schr\"{o}dinger equation given by 
\begin{equation}
i\hbar\frac{\partial  \Psi (\bf {k}, \bf {t})}{\partial t} =  \hat{H}^{T}_\lambda(t) \Psi (\bf {k}, \bf {t}) .
\label{SCHE}
\end{equation}

Based on the Floquet's theorem, the solution to the Eq.\,\eqref{SCHE} satisfies the periodic condition in time,
\begin{equation} 
 \Psi (\bf{k}, \bf{t+T})=\text{e}^{-i \hat{{\cal H}}_{\lambda,\zeta }^{Tot}T/ \hbar} \Psi (\bf{k}, \bf{t}) .
 \label{PSI}
 \end{equation}
 where $\hat{{\cal H}}_{\lambda,\zeta }^{Tot}$ is the Floquet effective Hamiltonian. We obtain this Hamiltonian by employing the Floquet-Magnus expansion for a high frequency periodic field. In this expansion, the effective Hamiltonian is given by,

\begin{equation}
\hat{{\cal H}}_{\lambda,\zeta }^{Tot} = \sum_{n=0} ^{\infty}  H^{(n)}_{eff}
\label{EFF}
\end{equation}
where $H^{(n)}_{eff} \sim \Omega^{-n}$ for 1, 2, ... and

\begin{equation}
 H^{(0)}_{eff}= \frac{1}{T} \int_{0}^{T} \hat{H}^{T}_\lambda(t) dt = \hat{{\cal H}}_\lambda(k_x,k_y)
\label{EFF1}
\end{equation}

\begin{equation}
H^{(1)}_{eff}= \frac{1}{2 i \hbar T} \int_{0}^{T} dt_{1}\int_{0}^{t_{1}} dt_{2} \left[\hat{H}^{T}_\lambda(t_{1}), \hat{H}^{T}_\lambda(t_{2})\right]
\label{EFF2}
\end{equation}

\begin{equation}
H^{(2)}_{eff}= \frac{1}{3! (i \hbar )^2 T} \int_{0}^{T} dt_{1}\int_{0}^{t_{1}} dt_{2}\int_{0}^{t_{2}} dt_{3} \left(\left[\hat{H}^{T}_\lambda(t_{1}),\left[\hat{H}^{T}_\lambda(t_{2}), \hat{H}^{T}_\lambda(t_{3})\right]\right] + \left[\hat{H}^{T}_\lambda(t_{3}),\left[\hat{H}^{T}_\lambda(t_{2}), \hat{H}^{T}_\lambda(t_{1})\right]\right] \right) .
\label{EFF3}
\end{equation}

The leading-order correction is the first-order correction in the series expansion. This  is given as

 \medskip
\par

 \begin{eqnarray}
 \label{eq2}
H^{(1)}_{eff}= \left(
\begin{array}{cccc}
0& v_{1}v_{2}\zeta & v_{1}v_{+}\zeta \lambda& 0\\
v_{1}v_{2}\zeta  & 0  &  0 &v_{1}v_{+}\zeta \lambda\\
v_{1}v_{+}\zeta \lambda & 0&0 &-v_{1}v_{2}\zeta \\
0&v_{1}v_{+}\zeta \lambda &-v_{1}v_{2}\zeta & 0
\end{array}
\right)  .
\end{eqnarray} 

The total effective Hamiltonian after expansion is

\begin{eqnarray}
\label{eq3} 
&& \hat{{\cal H}}_{\lambda,\zeta }^{Tot} =  \hat{{\cal H}}_{\lambda}(k_x,k_y) + H^{(1)}_{eff} \\
\nonumber 
&=& \left(
\begin{array}{cccc}
k_{y}(-v_{-}\lambda- v_{+}\lambda)& v_{1}v_{2}\zeta & -i k_{x}v_{1}+\alpha \Delta+v_{1}v_{+}\zeta\lambda& k_{y}v_{2}+\Delta\lambda\\
v_{1}v_{2}\zeta  &k_{y}(-v_{-}\lambda- v_{+}\lambda) &k_{y}v_{2}+\Delta\lambda&-i k_{x}v_{1}+\alpha \Delta+v_{1}v_{+}\zeta\lambda\\
i k_{x}v_{1}+\alpha \Delta+v_{1}v_{+}\zeta\lambda & k_{y}v_{2}+\Delta\lambda&k_{y}(-v_{-}\lambda+ v_{+}\lambda)&-v_{1}v_{2}\zeta \\
k_{y}v_{2}+\Delta\lambda&i k_{x}v_{1}+\alpha \Delta+v_{1}v_{+}\zeta\lambda&-v_{1}v_{2}\zeta & k_{y}(-v_{-}\lambda+ v_{+}\lambda)
\end{array}
\right)
\end{eqnarray}

where the wave vectors, velocities and energies are assumed dimensionless. We shall reintroduce their dimensions for our numerical calculations. 

\medskip
\par

In this Floquet Magnus expansion the off-resonant frequency is considered to be higher than both energy of the charge carrier in the system and light electron interaction energy. i.e, higher than any other energy scale employed by the system. In our numerical calculations, we have assumed that the illumination frequency is in the THZ range ($\sim 1.5\times 10^{14}$ HZ ). At high frequency in the strong THZ range, we assume that  $ \upsilon_{F} e E_0/ \Omega << \hbar \Omega$, i.e., $\zeta <<1$. For $\zeta = 0.2$, and illumination energy $\sim $ 0.1 eV, the pumping intensity would be about 3000 KV/m. Since the low-energy model describe most of the properties of the 2D material, we believe that our model would also be able to accurately describe the optical properties of this material with this frequency range as it is true for graphene in Ref.[\onlinecite{SK18b}]. In this approximation, we do not consider the Floquet index $"n"$ to classify the different sidebands. This is because the quasienergy spectrum only in the first Brillouin time-zone [$-\Omega/2,\Omega/2$] corresponds to $n=0$ satisfy the condition that at ${\bf A}(t)\to 0$, the energy spectrum is the spectrum of 1$T^{\prime}$MoS$_2$ in the absence of irradiation. Moreover, the solutions belongings to the $n$th Brillouin zone in the Floquet spectrum are those of the first Brillouin zone shifted by $n$ units. This is why, the solution does not incorporate  couplings to the Floquet sidebands. 
\medskip
\par

The corresponding eigenvalues and eigenvectors for the circularly polarized 1T$^\prime$MoS$_2$ are given by these expressions.

\begin{eqnarray}
  \mathbb{E}_{\xi,s,\lambda}^{Tot}({\bf k})= \mathbb{E}_{\xi,s,\lambda}^{Tot}(k_x,k_y)&=& -\lambda \upsilon_{-} k_{y}+ \xi \sqrt { k_{x}^{2}\upsilon_{1}^{2} +  k_{y}^{2}(\upsilon_{2}^{2}+\upsilon_{+}^{2})+ \upsilon_{1}^{2}\upsilon_{2}^{2}\zeta^{2}+2 \  k_{y}v_{2} \Delta (- s \alpha+\lambda)+(\alpha\Delta- s \Delta\lambda+\upsilon_{1}\upsilon_{+}\zeta\lambda)^{2}}
\nonumber\\
    & = &-\lambda \upsilon_{-} \  k_{y}+ \xi \sqrt{|{\cal \tilde{D}}_{s}^{\lambda}|^{2} + \left( s\  \upsilon_{+} k_{y} + \upsilon_{1}\upsilon_{2}\zeta \lambda\right)^2} \   ,
\label{ANSWER}
\end{eqnarray}
where  we have introduced  ${\cal \tilde{D}}_{s}^{\lambda}\equiv   i   v_{1} k_{x}  - s(  \upsilon_{2} k_{y} + \lambda \Delta) +\alpha \Delta + v_{1}v_{+} \lambda \zeta$.

\medskip
\par

We can write in closed form for the eigenvector as

\begin{equation}
\psi_{\xi,s}^{\lambda}  ({\bf k},{\bf r}) =\left(
\begin{array}{c}
\dfrac{ s v_{1}v_{2}\zeta + \lambda   k_{y}( v_{+} -v_{-}) -   \mathbb{E}_{\xi,s,\lambda}^{Tot}}{s {\cal \tilde{D}}_{s}^{\lambda}} \\[12pt]
\dfrac{s v_{1}v_{2}\zeta + \lambda   k_{y}( v_{+} -v_{-}) -   \mathbb{E}_{\xi,s,\lambda}^{Tot}}{-{\cal \tilde{D}}_{s}^{\lambda}}) \\[12pt]
-s\\[12pt]
1
\end{array} \right)
\frac{e^{i{\bf k}\cdot {\bf r}}}{\sqrt{A}}, 
\end{equation}

\newpage

\begin{figure}[!ht]
\centering
\includegraphics[width=0.7\linewidth]{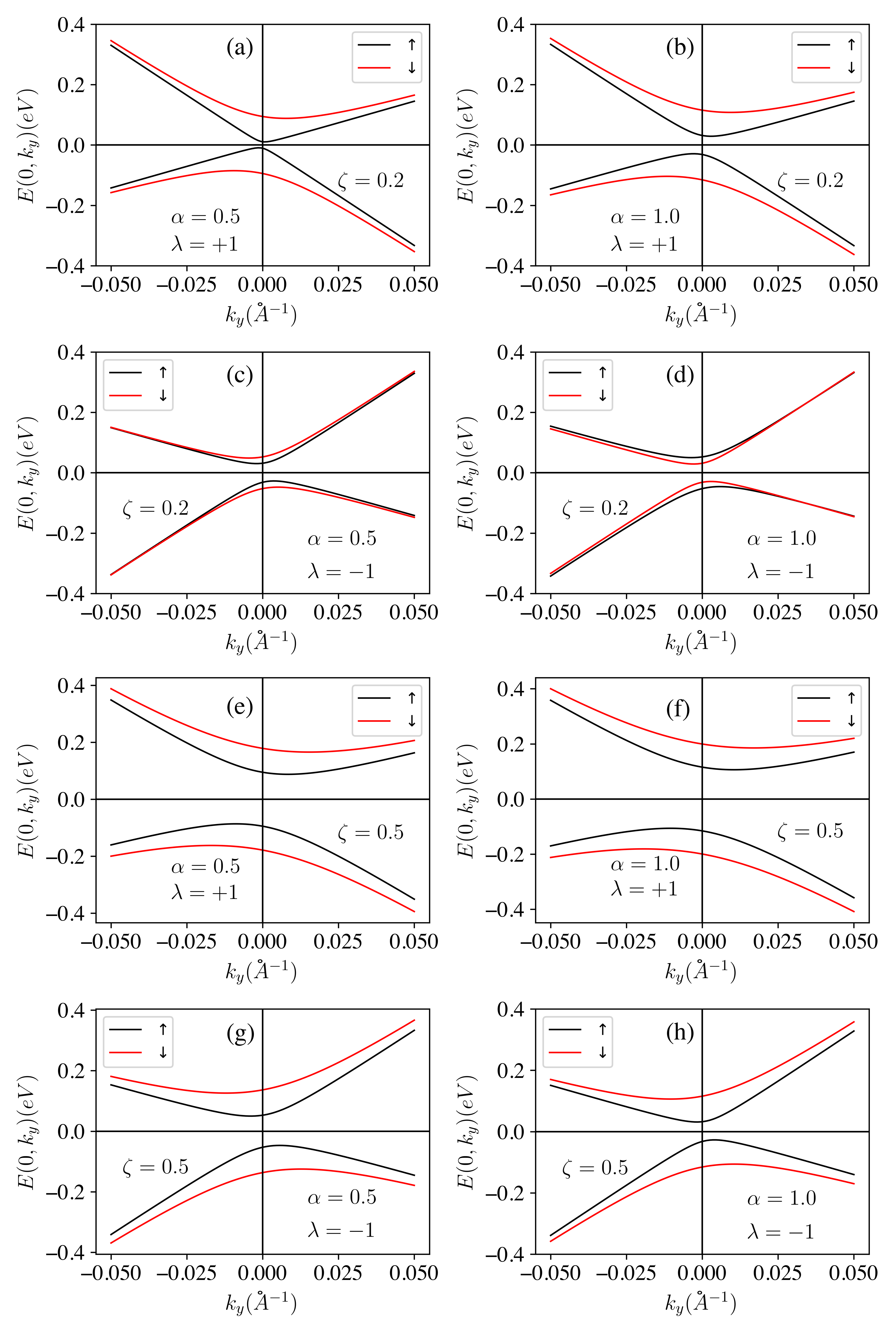}
\caption{ (Color online) Floquet engineered valence and conduction energy bands in tilted 1T$^{\prime}$MoS$_2$ monolayer in the presence of a vertical electric field. The energy of the dressed states  is plotted as functions of $k_y$ when $k_x=0$. In the plots (a),(b),(c) and (d) in the upper panel, we chose  the electron-light interaction energy parameter $\zeta =$ 0.2,  whereas we chose $\zeta$ = 0.5 for the other plots (e),(f),(g) and (h) in the lower panel. The vertical electric field is equal to half the critical value, (i.e.,  $\alpha$ = 0.5) in (a),(c), (e) and (g);  and equal to  the critical value, (i.e.,  $\alpha$ = 1) in (b),(d),(f) and (h). The dispersions at both Dirac points correspond to $\lambda = \pm1$ are presented as indicated in the figure.}
\label{energy_2}
\end{figure}

\medskip
\par

Our numerical results for the energy dispersion of tilted 1T$^{\prime}$MoS$_2$ in the presence of circularly polarized light are presented in Fig.\ \ref{energy_2} for two coupling strengths, i.e., $\zeta =$ 0.2 and  0.5, at both Dirac points i.e., $\lambda = \pm 1$, and for two values of the vertical electric fields i.e., $\alpha =$ 0.5 and 1.0. The additional term in the energy dispersion due to irradiation is \( \sqrt {( \upsilon_{1}^{2}\upsilon_{+}^{2}\lambda^2+  \upsilon_{1}^{2}\upsilon_{2}^{2})\zeta^{2}+ 2\upsilon_{1}\upsilon_{+}\zeta\lambda\Delta (\alpha - s \lambda)}\) which is zero for $\zeta=0 $. It shows that circularly polarized light  in 1T$^{\prime}$MoS$_2$ does not always widen the direct band gap at $|{\bf k}|= 0$ as in graphene. It may reduces the gap depending on the different parameters such as $\alpha, \zeta, \lambda$ and spin-orbit coupling $\Delta$ . For example, at the positive Dirac point ($\lambda =$+1) in (a)  in Fig. \ref{energy_2},  for $\alpha =$ 0.5, the already existing small gap in the absence of irradiation is reduced in up spin channel and is increased by a small amount in down spin channel. However for $\alpha =$1.0 in (b) up spin channel also opens the gap and the already existing gap for down spin channel becomes wide when the states are irradiated with low  interaction energy corresponding to $\zeta =$0.2. Similarly, for the same value of $\zeta$, at the negative Dirac point ($\lambda = -1$), the up spin gap is further reduced  and the down spin gap increased for $\alpha = $0.5 while a small gap is opened for down spin channel and up spin gap is further reduced for $\alpha = $1.0 which can be seen from (c) and (d) of the same figure.  The outcome is different when we increase the irradiation parameter $\zeta$ equal to $0.5$. Although the width of the gap may vary depending on different  chosen parameters, but the gap is always open as shown in (e),(f), (g) and (h) of Fig. \ref{energy_2}. The gap reshuffling is for the bands around  $|{\bf k}|= 0$. The energy dispersion for finite $|{\bf k}|$  always remains the same.  The bands are still tilted along the $k_y$ axis and remains un-tilted along $k_x$. The interesting thing is that for dressed states, the spin subbands are not reversed when we go from one Dirac point to another as for undressed states.

\begin{figure}[ht]
\centering
\includegraphics[width=0.7\linewidth]{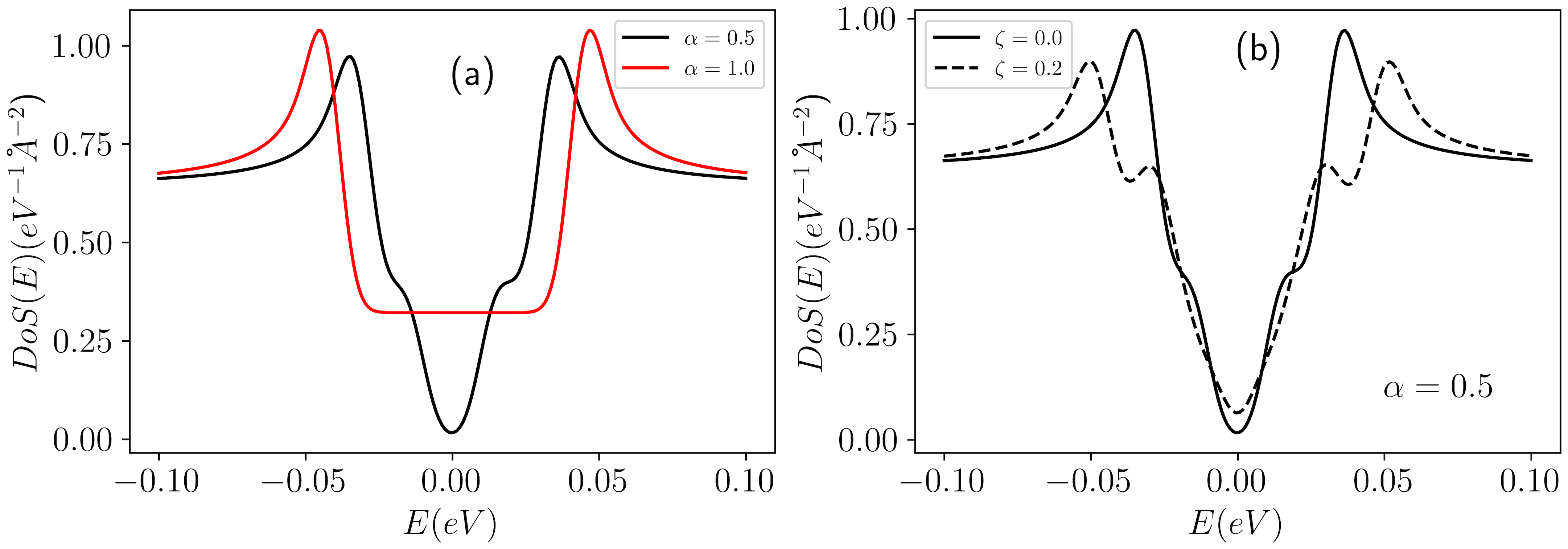}
\caption{(Color online) Density-of-states (DoS) for two values of the vertical electric field parameter $\alpha$ as indicated for undressed states (a), and  for vertical electric field half of the critical field for dressed states with  $\zeta = 0.2$  and undressed states $\zeta = 0.0$  in (b). }
\label{DoS}
\end{figure}

The effects due to the gaps between the energy subbands that are generated by  $\alpha $ and $\zeta$ are clearly reflected in the density of states.  The presented density of states in Fig.\ \ref{DoS} are calculated from  the equation

\begin{equation}
DoS(E)=  \sum_{\lambda=\pm1} \, \sum_{\zeta, s=\pm1} \int \frac{d^2{\bf k}}{(2\pi)^2} \delta (E -  \mathbb{E}_{\xi,s,\lambda, \zeta}({\bf k}))
\end{equation}
for two values of $\alpha$ for undressed states in (a) and for two values of $\zeta$ at $\alpha=$0.5 in (b).  In this expression  $\delta (x)$ is the Dirac delta function and $\mathbb{E}_{\xi,s,\lambda, \zeta}({\bf k})$ are energy dispersion. where $\zeta =$0 represents the dispersion for undressed states. For the numerical calculation we employ a Lorentzian representation for the delta  function i.e.,  \(\delta(x) \to \frac{\eta}{\pi} \frac{1}{\eta^2+x^2} \)  with a broadening of $\eta=0.003$. It is also important to note that contribution from all energy subbands including spin (s), valley ($\lambda$) and electron-hole ($\xi$) are added up to the DoS.

 \section{Polarization function, plasmon dispersion and decay rate}
\label{sec3}

\medskip
\par

Using linear response theory,  one can obtain    an expression for the induced  electron density  fluctuation due to an external perturbation.  This follows  from the equation of motion for the density matrix in the absence of dissipation.    This leads to  a result in terms of the external potential  $\phi({\bf r},\omega)$ and the density-density response function as follows.\cite{GG1}

\begin{equation}
\delta n({\bf  r},\omega)=e\int  d{\bf  r}^\prime \  \Pi^{(0)}( {\bf  r}, {\bf  r}^\prime)\phi_{ext}({\bf  r}^\prime ,\omega) \  ,
\end{equation}
for non-interacting particles,  where  the density-density response function $\Pi^{(0)}({\bf r}, {\bf r}^{\prime} , \omega)$ is given in terms of ``four” wave functions $\phi_\nu({\bf r})=\hat{\psi}({\bf  r})|\nu>$, by
\begin{equation}
\Pi^{(0)}( {\bf  r}, {\bf  r} ^{\prime} , \omega) =    \sum_{\nu , \nu^{\prime}}   \frac{ f_0 (\epsilon_{\nu^\prime} )- f_0 (\epsilon_{\nu }    )}{\hbar \omega - \epsilon_{\nu} + \epsilon_{\nu^{\prime}}} 
 \phi_{\nu}^{\ast}({\bf  r}^\prime) \phi_{\nu^\prime}({\bf  r}^\prime) \phi_{\nu^{\prime}}^\ast({\bf  r} ) \phi_{\nu }    ({\bf  r} )  \   .
\label{GG:2}
\end{equation}

If we now write \(|\nu>\Rightarrow |\nu , {\bf  k}>\frac{e^{i{\bf  k}\cdot{\bf  r}}}{\sqrt{A}} \) , where     $A$ is a normalization area and then substitute into Eq. \ (\ref{GG:2}), we immediately see that $\Pi^{(0)}( {\bf  r}, {\bf  r} ^{\prime} , \omega) $  is a function of the difference ${\bf r}-{\bf  r}^\prime$ and does not depend separately on the two spatial coordinate variables.   Finally, taking the Fourier transform with respect to  ${\bf  r} - {\bf  r} ^{\prime}$ , we obtain the expression for the polarization function as a function of wave vector and frequency $\Pi^{(0)}({\bf q},\omega)$ which includes the form factor $|<\nu^\prime,{\bf  k}+{\bf  q}|\nu,{\bf  k}>|^2$. The polarizability of the medium is also depend on the  chemical potential $\mu$ and the valley index    $\lambda$ which were suppressed in this brief derivation.

\begin{equation}
\Pi^{(0)}_\lambda({\bf q},\omega;\mu) =  \frac{1}{4 \pi^2}  \int d^2 {\bf k} \sum\limits_{s, s' = \pm 1}
\sum\limits_{\xi, \xi' = \pm 1}
{\cal F}_{s,\xi,\lambda;s',\xi',\lambda}({\bf k},{\bf q})\frac{f_0(       E_{\xi,s}^{\lambda}({\bf k})    -\mu)-f_0(
   E_{\xi',s'}^{\lambda}({\bf k}+{\bf q})-\mu)}
{    E_{\xi,s}^{\lambda}({\bf k})   -
   E_{\xi',s'}^{\lambda}({\bf k} +{\bf q})
   - \hbar (\omega + i \delta^+)}\ .
\label{pimain}
\end{equation}
Here, the momentum transfer ${\bf q}$ is measured from the K point within a valley, $f_0(E-\mu)=\{1+\exp[(E-\mu)/k_BT]\}^{-1}$ is the Fermi-Dirac distribution functions, $E_{\xi,s}^{\lambda}({\bf k})$ is given by Eq.\,  (\ref{ANSWER}),  $\mu$ is the chemical potential of the system, $T$ is the temperature, and the overlap function is 

\begin{equation}
 {\cal F}_{s,\xi,\lambda;s',\xi',\lambda}({\bf k},{\bf q})   = \left|<s,\xi,\lambda, {\bf k}|s',\xi',\lambda,{\bf k}+{\bf q}> \right|^2.
\label{FS}
\end{equation}
The two Fermi-Dirac distribution functions appearing in the numerator of Eq.\  (\ref{pimain}) can be replaced by two Heaviside unit step functions $\Theta(\mu - E_{\zeta^{\prime}, s^{\prime}}^{\lambda}(k+q))$ - $\Theta(\mu - \mathbb{E}_{\zeta, s}^{\lambda}(k))$  at T = 0 K. There is a small positive imaginary part added to the frequency, i.e., $\omega\to \omega+i\delta^+$ in the denominator, which gives the decoherence effect or natural damping of the plasmons. For decoherence,  we must choose $\delta^{+}\ll \omega_p$, the plasmon frequency.  In this representation, we can easily separate the polarizability $\Pi^{(0)}_\lambda({\bf q},\omega;\mu)$ into the polarizability when the Fermi level just lies below the lower of the conduction band, i.e., $\mu < G(\alpha,\lambda,\xi)$.  The  second contribution to the polarizability exists when the Fermi level rises up to the second conduction subband and we have

\begin{equation}
\Pi^{(0)}_\lambda({\bf q},\omega;\mu) =  \Pi^{(\infty)}_\lambda({\bf q},\omega) + \Theta(|\mu|- G(\alpha,\lambda,\xi)) \Pi^{(\mu)}_\lambda({\bf q},\omega) .
\label{pseparate}
\end{equation}
In this notation,  $G(\alpha,\lambda,\xi)$ represents the energy difference between the minimum of the conduction band and the maximum of the valance band, i.e., $| Min(\mathbb{E}_{+})- Max(\mathbb{E}_{-})|/2$  for chosen values of $\alpha, \lambda$ and $\xi$.

\medskip
\par

The various transitions between occupied and unoccupied electron states contribute to the total polarizability.  As we further explain below, the selection rule for transitions between  eigenstates is governed by the form factor involving the eigenstates as defined in Eq.\ (\ref{FS}).  The two Fermi-Dirac distribution functions ensure that only transitions involving electron-hole pairs can possibly  make a finite contribution to the polarizability. For 1T$^{\prime}$ MoS$_2$,  the transition between two states $|\lambda,\xi,s,{\bf k}>$   and $|\lambda,\xi^\prime,s^\prime,{\bf k}+{\bf q}>$ is possible not only when the initial state is occupied and the final state is empty but  requires the overlap of their wave functions is finite. That is, $  E_{\zeta^{\prime}, s^{\prime}}^{\lambda}(k+q))$ - $ E_{\zeta, s}^{\lambda}(k)$ $\neq$0 and ${\cal F}_{s,s'}({\bf k},{\bf q})  \neq 0$. There is a strong overlap between the wave functions with the same spin, thereby making a larger contribution  to the polarizability. However, the wave functions corresponding to those states having  opposite spin directions, overlap poorly and the transition between these two states could be neglected due to the selection rule governed by the overlap function. The polarizability from the first term on the right side of Eq. \  (\ref{pseparate}), is purely the interband polarizability. This is true when $|\mu|< G$ and is independent of $\mu$.  When the  Fermi level is increased and at least one conduction subband is occupied, i.e., ($\mu > G $), transitions are allowed between the subbands, thereby  making finite contributions to the polarizability. The second term on  the right side of Eq. \ (\ref{pseparate}) gives  both interband and intraband corrections to the polarizability when $\mu > G $. We note that  when integrating over the states, we exclude those states below the minimum of the conduction band by introducing an extra step function  $\Theta(|\mathbb{E}|- G(\alpha,\lambda,\xi))$ as it is  also done for gapped graphene in Ref.\  [\onlinecite{SK21}].

The polarization function $\Pi^{(0)}_\lambda({\bf q},\omega ;\mu)$ presented in Eq.\,  (\ref{pimain}) is a crucial quantity for determining the  collective properties of the host material.    The plasmon dispersions can be obtained from the poles of  $\Pi^{RPA}  ({\bf q},\omega) =\Pi  ^{(0)}  ({\bf q},\omega )/ \epsilon({\bf q},\omega) $ where $\Pi  ^{(0)}  ({\bf q},\omega )=\sum\limits_{\lambda=\pm 1 } \Pi^{(0)}_\lambda({\bf q},\omega ;\mu)$ and $\epsilon({\bf q},\omega)$ is the dielectric function which is given by

\begin{equation}
\epsilon({\bf q},\omega) = 1 - V(q)  \Pi^{(0)}  ({\bf q},\omega ) \
\label{eps}
\end{equation}
Also, in this notation, $V(q)=e^2 /(2\epsilon_0\epsilon_bq)$ is the 2D Coulomb interaction and $\epsilon_b$ is the dielectric constant of the host material.

\begin{figure}[ht]
\centering
\includegraphics[width=0.45\linewidth]{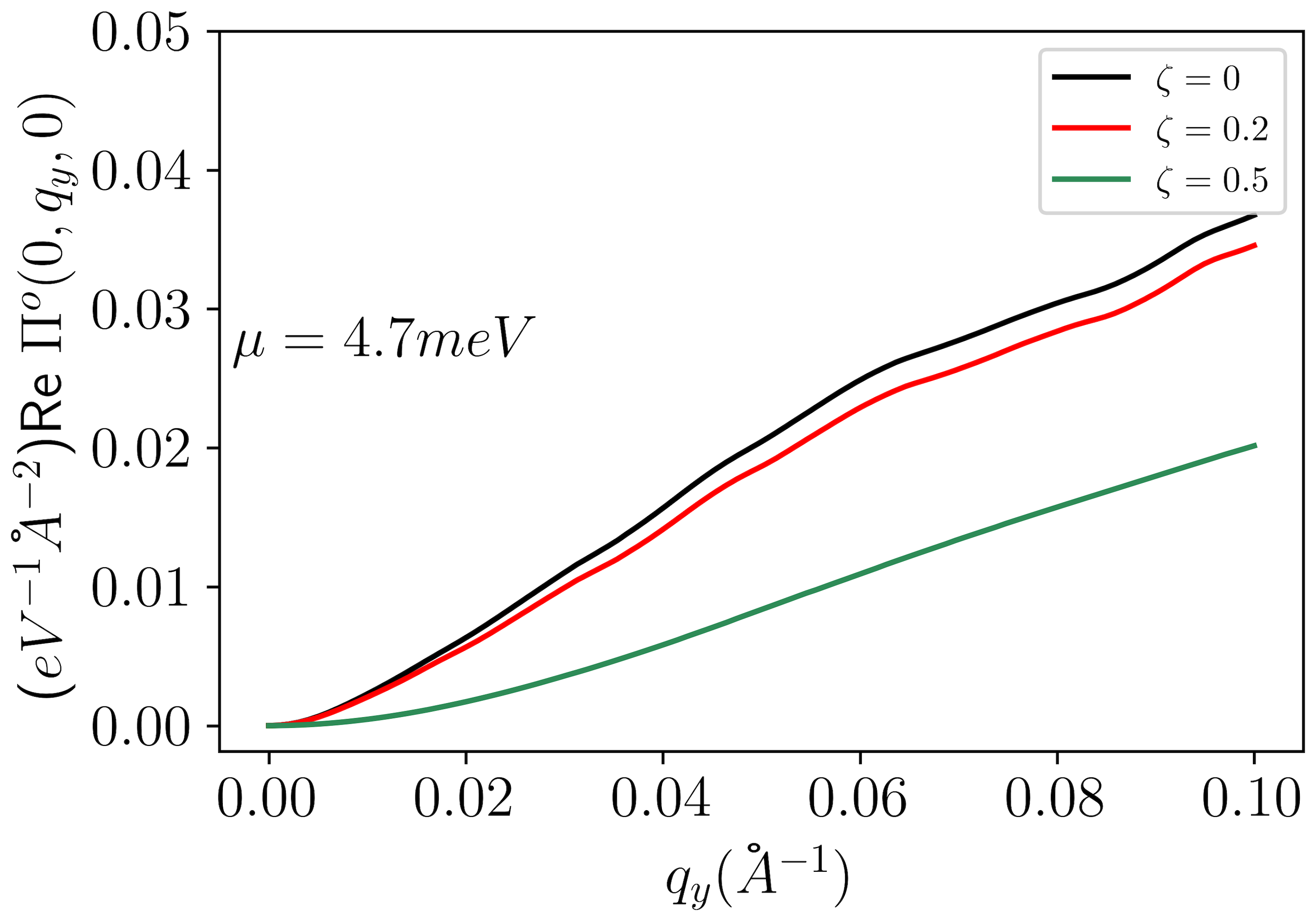}
\caption{(Color online)   Two-dimensional line plots of  the noninteracting static polarization function $\Pi^{(0)}({\bf q},\omega=0;\mu)$ along $q_y$ for  perpendicular electric field $\alpha = 0.5$. This is an interband contribution for chemical potential $\mu$ less than the band gap, i.e., the Fermi level lies below the minimum of the conduction band. The different curves are for three chosen values of $\zeta $ as indicated. }
\label{pol_1}
\end{figure}

\begin{figure}[ht]
\centering
\includegraphics[width=0.7\linewidth]{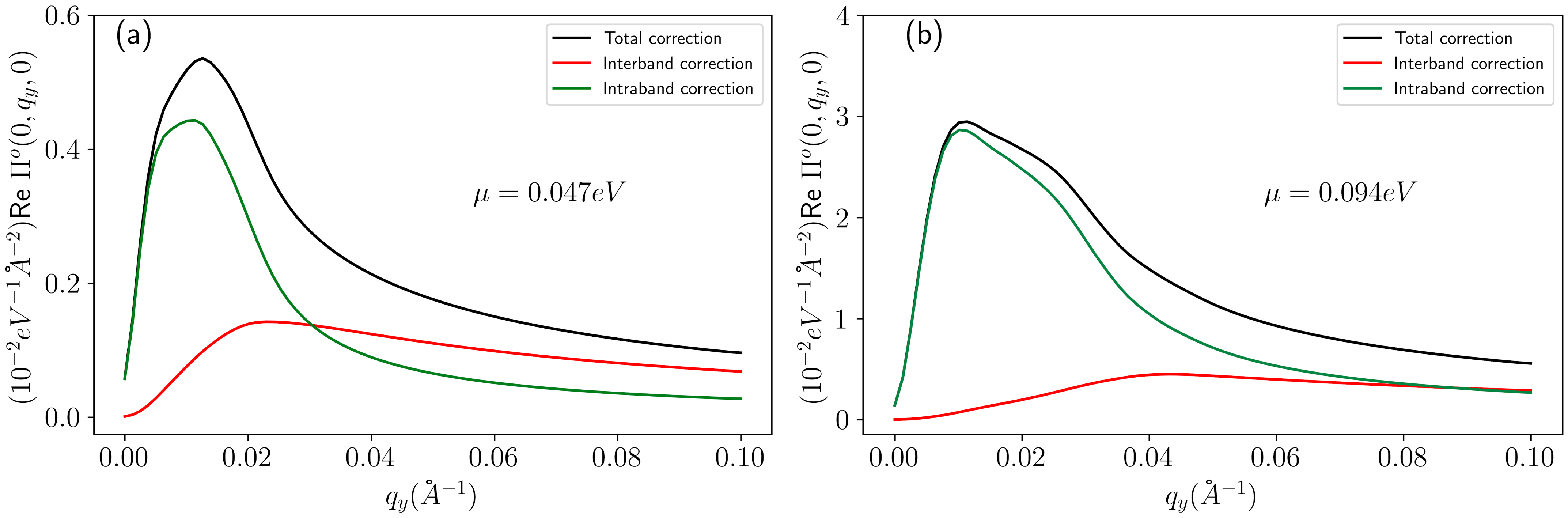}
\caption{ (Color online)   Two-dimensional line plots of  the correction-term to the noninteracting static polarization function $\Pi^{(0)}({\bf q},\omega=0;\mu)$  when the chemical potential  $\mu$ is greater  than the band gap. i.e.,  the lower conduction band is occupied but the upper conduction band is not in (a). Both conduction bands are occupied in (b). We separated the interband and intraband corrections. The  perpendicular electric field $\alpha$ is chosen to be half the critical field and in the absence of irradiation.   The graphs are  qualitatively  similar in nature in the presence of irradiation.}
\label{pol_1_corr}
\end{figure}

\begin{figure}[ht]
\centering
\includegraphics[width=0.7\linewidth]{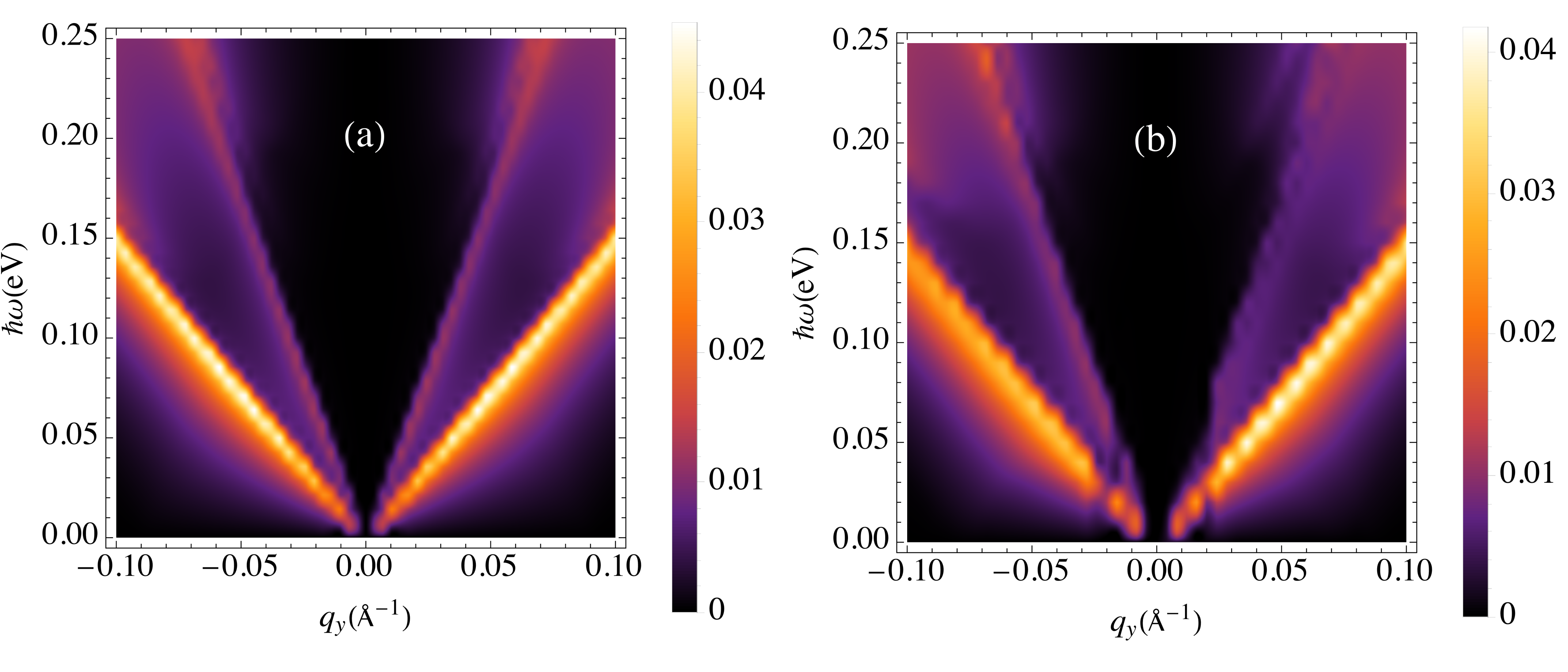}
\caption{ (Color online)   Density plots of Im $\Pi^{(0)}({\bf q},\omega;\mu)$ . The shaded non zero area  is the single-particle excitation region where the plasmons are Landau damped. Plot  (a)  is without irradiation (i.e.,  $\zeta =$ 0) and plots  (b) is in the presence of irradiation with   $\zeta =$ 0.2, along the $q_y$  direction for  $q_x = $ 0. }
\label{pol_2}
\end{figure}

\begin{figure}[ht]
\centering
\includegraphics[width=0.8\linewidth]{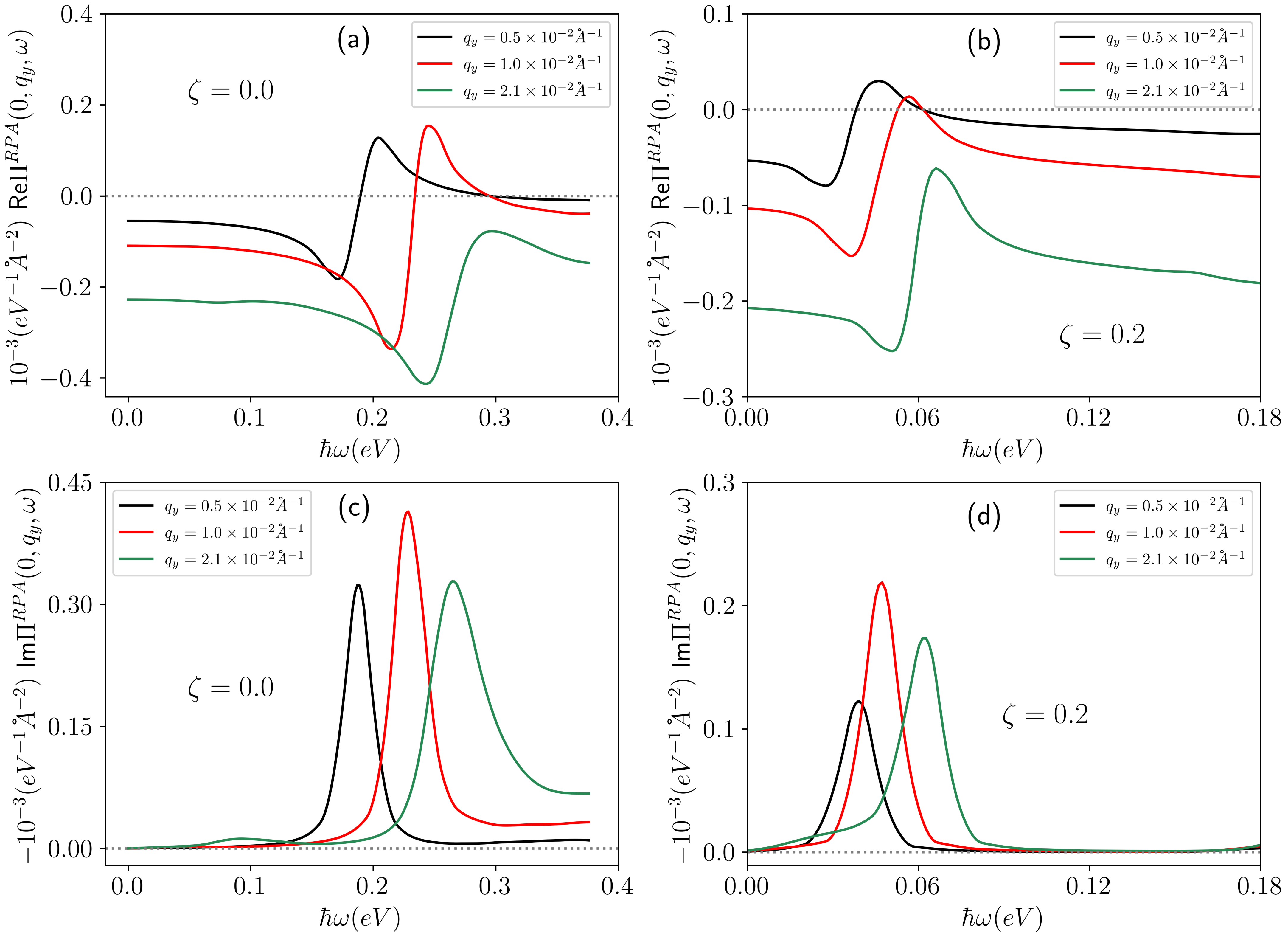}
\caption{ (Color online) Real and imaginary parts of $\Pi^{(RPA)}({\bf q},\omega;\mu)=\Pi^{(0)}({\bf q},\omega;\mu)/\epsilon({\bf q},\omega;\mu)$ as functions of $\hbar \omega$ for chosen $q$. The left panel (a) and (c) are without irradiation. The right panels (b) and (d)  are for an irradiation field  being present   with    $\zeta =$ 0.2. Plots are along the $q_y$ directions with $q_x =$0 and for $\mu =$0.094eV.  The resonance in the real part developed for some values of $q_y$ and which signals the existence of a collective mode in this region of $q_y$ and $\omega$.  }
\label{pol_rpa}
\end{figure}

\medskip
\par
The  function $\Pi^{(0)}_\lambda({\bf q},\omega ;\mu)$ is a complex function of both frequency $\omega$ and wave vector $\bf{q}$. Additionally, it depends on the level of doping, i.e., the chemical potential $\mu$. In the limit $\omega \to 0$, the polarization has its own importance. The real part of  the static polarization $\Pi^{(0)}(q,\omega=0)$ decides the screened potential of charged impurities in the sample. On the other hand, due to the fact that  the imaginary part of  $\Pi^{(0)}(q,\omega)$ is  antisymmetric, its value vanishes at zero frequency. Our numerical results for the static polarization function at chemical potential $\mu=4.7meV < G(\alpha,\lambda,\xi)$ and $\alpha = 0.5$ for both non-irradiated and irradiated 1T$^{\prime}$MoS$_2$ for two values of $\zeta$ (0.2 and 0.5) are plotted in Fig. \ref{pol_1}.  The polarization is  linear along $q_{y}$ for $q_{x} =$0 and is purely due to interband transitions. The effect of dressed states with circularly polarized light on the static polarization is not much significant for  very small interaction energy  $\zeta \sim 0.2$, but the polarization is  suppressed when the interaction is strong enough but $\zeta$ is still less than one. However in Fig. \ref{pol_1_corr}, we have plotted the correction to the polarization due to states $\mu > G(\alpha,\lambda,\xi)$ separating the contributions from interband and intraband transitions. It is seen that the correction has delta like peak in long wavelength region and is saturates to a minimum value as $q$ rises. The correction is enhanced when both conduction subbands are occupied at $\mu = 0.094 eV$ in (b) than that at $\mu = 0.047eV$ in (a) where only the lower conduction band is occupied. The total polarization is always linear as the interband transition dominates the polarization. In contrast, monolayer graphene has constant static polarization for $q< 2k_F$ and increases linearly for $q>2k_F$ \cite{fengping}.

\medskip
\par

There are various uses for the dynamical polarization $\Pi^{(0)}({\bf q},\omega)$. It helps us investigate the response of an electronic system to an external probe. Its real part is employed in determining the plasmon branches while the imaginary part gives the region in $(q,\omega)$ space where plasmons are Landau damped. Here, we are interested in the polarization of the system in the presence of a  circularly polarized light field. In the presence of the electromagnetic field, the analytical behavior of the polarization function remains the  same, i.e., is analytic in the upper half-plane. The effect of a dressing field on the sample is to renormalize the Fermi velocity and redistribute  the spectral weight of the transitions. For undoped ($\mu =$0) or doped 1T$^{\prime}$MoS$_2$ with $\mu< G$, the imaginary  part of $\Pi^{(0)}$ is zero. However, for $\mu> G$ , it is finite for real $\omega$. Therefore, for finite $\mu> G$, the plasmon excitations could be  Landau damped by intraband single-particle excitations.  Our numerical results for the imaginary parts of the non-interacting polarization function  at $T = 0$ K for $\mu = 0.094eV$ is presented in Fig.\  \ref{pol_2}. These results depict the regions where the plasmon modes are excited and single-particle excitations are located. The ``v-shape” shaded region along $q_y$ in (a) and (b) of Fig.\  \ref{pol_2} corresponds to the non-zero values of the imaginary part. Plasmons are Landau damped when they  lie in this region. They remain long lived  only when they are  present in the non zero region of the real part of $\Pi^{(0)}$. In the presence of irradiation for chosen $\mu$, even the single particle excitations are suppressed as there are not enough electrons to be excited to the conduction band. Therefore, the  intensity of the polarization in the low $q$ region  in (b)  is reduced as the polarization in this region is due to intraband transitions which is reduced due to the presence of irradiation. In generating  these results, we have added the contributions from both Dirac points.

\medskip 
\par
For the interacting system, the low-Coulomb excitations are well described within the  random phase approximation. In Fig.\  \ref{pol_rpa}, the real and imaginary parts of $\Pi^{(RPA)}({\bf q},\omega;\mu)$ are plotted. In the absence of irradiation in Fig.\  \ref{pol_rpa}(a), it is seen that  the real part has resonances for  $q_y= 0.005 \AA^{-1}$ and $0.010\AA^{-1}$, where the imaginary part has no damping in (c) leading to the existence of long lived plasmons. However, for larger $q_y$, the real part has not developed resonances and significant excitations due to the finite imaginary part. Our results show that plasmons are Landau damped in the low $q_y$ region and these plasmons decay into single particle excitations. Plasmons are also damped in the high $q_y$ region where they disappear after excitation. When a circularly polarized light field is applied to the system, the real part of the polarization acquires a resonance for large values of the wave vector and its range is increased as we increase the interaction energy. However, the real part becomes unstable for some values of $q_{y}$.  This is clear from Fig.\  \ref{pol_rpa}(b) which is for $\zeta = 0.2$.  Therefore, for dressed states, the self-sustained collective oscillations exist for shorter wavelength in comparison with that  for undressed states which may have varied slope. The situation will be different for finite temperature, which we will discuss in Sec.\  \ref{sec4}.

\medskip 
\par
The nonlocal plasmon-decay rate $\gamma[q,\omega_p(q)]$, close to the plasmon frequency $\omega\approx\omega_p(q)$, can be approximated as \cite{SK38,malcolm}

\begin{equation}
\label{gammap}
\gamma[q,\omega_p(q)]= \frac{
\text{Im}\left[
\Pi^{(0)} (q,\omega_p ) \right]}{
\text{Re} \left[
\displaystyle{\left.\frac{\partial\Pi^{(0)} (q,\omega  )}{\partial \omega}
\right|_{\omega = \omega_p}}
\right]}\ ,
\end{equation}
where $q$ is the wave number of the plasmon mode and $\omega_p(q)$ represents the dispersion of plasmon energy.  The magnitude of plasmon damping can be affected by three factors. The first and foremost is the nonzero $\text{Im}\left[\Pi^{(0)} (q,\omega_p ) \right]$ or the numerator in Eq.\,\eqref{gammap}, which determines whether the damping does or does not exist. $\gamma[q,\omega_p(q)]$ also depends on $\omega$ in $\text{Re}\left[\Pi^{(0)} (q,\omega)\right]$ or the denominator in Eq.\,\eqref{gammap}. 
To excite a plasmon mode, we need $\text{Re}\left[\Pi^{(0)} (q,\omega \ \, )\right]>0$, and meanwhile, we also require a steep change in this function with respect to $\omega$ around $\omega=\omega_p(q)$ so as to reduce the decay rate in Eq.\,\eqref{gammap}. Finally, the separation of the $\omega_p(q)$ region away from the damping region ensures a fully undamped damping-free plasmon mode.

  \medskip
\par

\begin{figure}[!ht]
\centering
\includegraphics[width=0.8\linewidth]{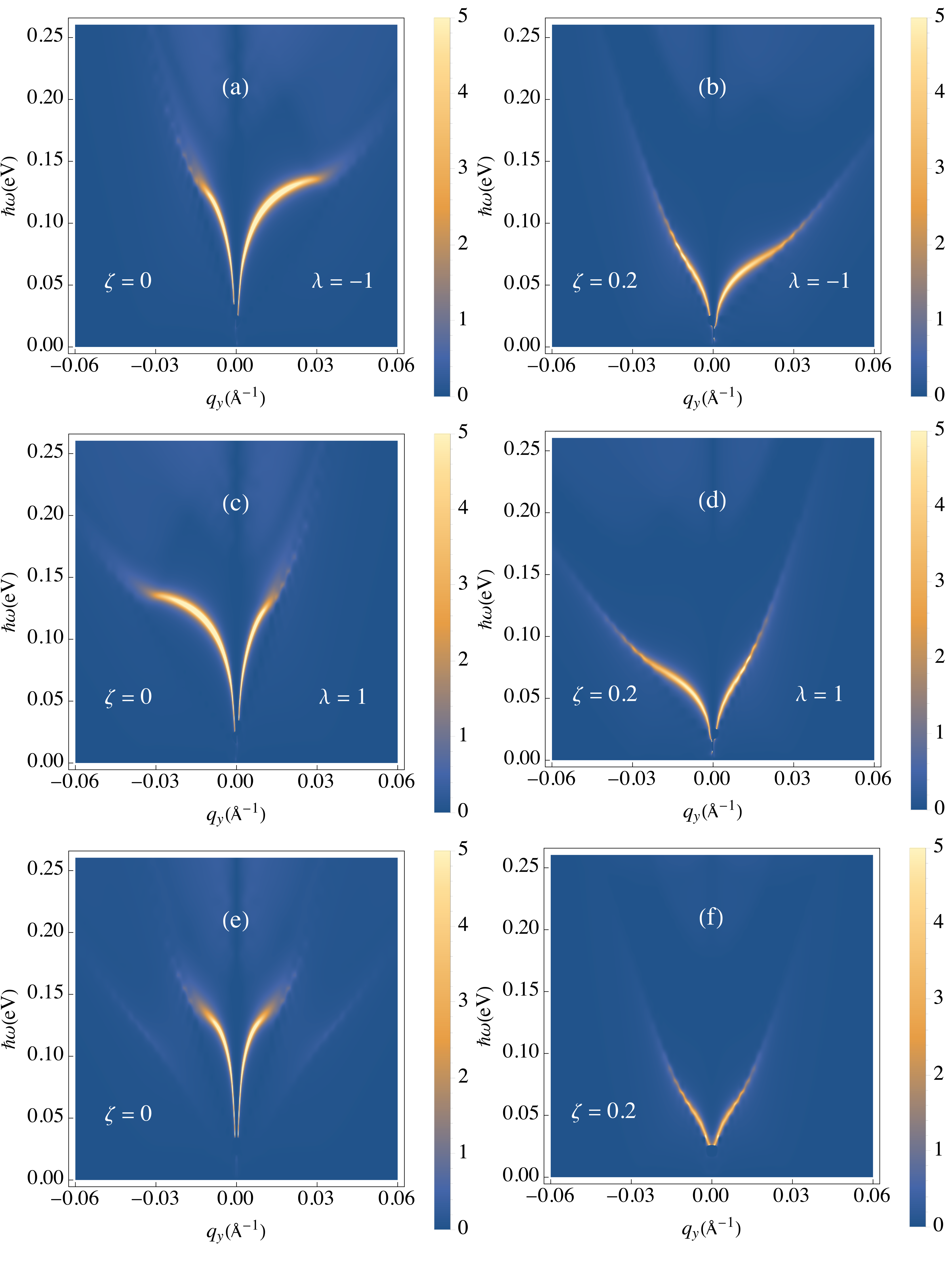}
\caption{ (Color online)  Density plots for the plasmon dispersion relation obtained from the loss function $\text{Im}[1/\epsilon(q, \omega)]$  as a functions of wave vector component $q_y$ when $q_x = 0$  and frequency $\omega$ in $1T^\prime$MoS$_2$  separating the contributions from two opposite Dirac points at T=0 K.  Plots (a),  (b) and (c), (d) are at specific Dirac points as indicated while plots (e) and (f) are when the polarization from both Dirac points are added up. The left panel is in the absence of irradiation, i.e., $\zeta =$ 0, whereas the  right  panel  is with irradiation (i.e., $\zeta =$ 0.2). The chemical potential $\mu$ is chosen to be 0.094eV. Note the Landau damping of the plasmon modes in the long wavelength limit showing that plasmons emerge from the particle-hole region at finite wave vector.   }
\label{loss_fn}
\end{figure}

\begin{figure}[ht]
\centering
\includegraphics[width=0.7\linewidth]{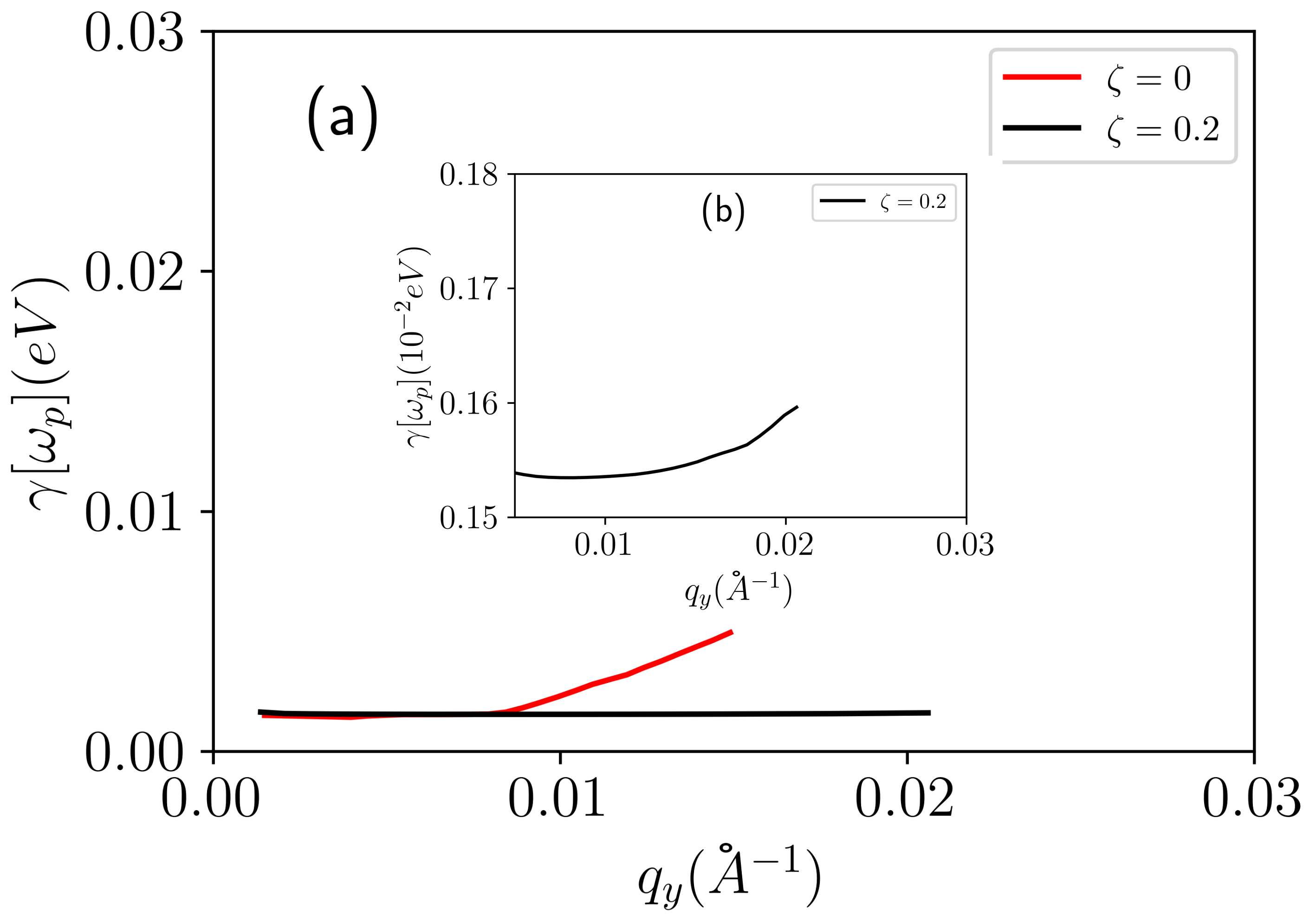}
\caption{ (Color online)  Plasmon decay rate due to Landau damping along $q_y$ (a) and a small part of the black curve in a chosen region of $q_{y}$ where damping starts  is shown in (b). The red curve is for  the case without irradiation and the black curve is with irradiation. The plasmon damping rate is reduced in the presence of irradiation. }
\label{damp}
\end{figure}

\medskip
\par
Our investigation into the dielectric function enables us to understand the plasmon dispersion for the chosen material. The zeros of this function in Eq.\ (\ref{eps}) correspond to the plasma modes. Equivalently, the peaks in the loss function Im$[1/\epsilon({\bf q},\omega)]$ yield the frequencies and intensities of the plasmon excitations.  In Figs.\  \ref{loss_fn}  and \ref{damp}, we present 2D density plots  of the plasmon dispersion relation obtained from the loss function as well as their damping rates. The dispersion at one Dirac point is the mirror reflection of other at $\bf{q} = 0$ as the dispersion in panel (a) and (b) at $\lambda = -1$ and dispersion in panel (c) and (d) at $\lambda = +1$ seems as a mirror images reflected at $q_y = 0$.
In general, in the absence of irradiation, the dispersion follows the $\sqrt{q}$ behavior in the long wavelength limit as it does for graphene and the  2D electron gas as well as other 2D materials like silicene. Unlike graphene and the 2D electron gas, no plasmons are  seen to exist in the very low frequency and $q \to 0$ region because of the intraband polarizability. For  non-irradiated 1T$^\prime$MoS$_2$, the  plasmons are Landau  damped in the short wavelength limit due to single-particle  transitions between bands. However, as a result of the modification of the gap and the redistribution of the spectral weight for irradiated 1T$^\prime$MoS$_2$, the plasmon frequencies are highly diminished and their intensity is also reduced. From the results in the right panel of Fig.\  \ref{loss_fn}, we note that the effect due to irradiation is equal on both Dirac points. It is also observed that the threshold of frequency for plasmons to exist is less for an irradiated sample compared to that for an non-irradiated sample. Despite the reduction of the plasmon frequency due to irradiation, the lifetimes of the plasmons are extended and are sustainable in the shorter wavelength region. The 2D dispersions in Fig.\  \ref{loss_fn} agrees with the particle-hole mode region in Fig.\  \ref{pol_2}, for the  polarization plot in Fig.\  \ref{pol_rpa} and the plots for the decay rates in Fig.\  \ref{damp}. In Fig.\  \ref{damp}, there is no effect due to irradiation in long wavelength region as both curves for the decay rate overlap  but the effect is significant in the short wavelength region as they are separated by a sizeable gap. It is because in the long wavelength region ($\sim10^{-5}$m $- 10^{-7}$m),  the wavelength of the plasmon mode and the irradiation ($\sim10^{-6}$m) are within the range of each other and hence the interplay between them is not significant. However in the short wavelength region, the wavelength of the plasmon mode is much shorter than that of the irradiation and hence the interplay is significant. The long lived plasmons are suitable for application purposes. The highly unstable plasmons are observed for graphene in the low $q$ region in the presence of circularly polarized ac electric field than their counterpart in graphene with no field, which is due to the anti-crossing between Floquet sidebands. \cite{SK18b}

\medskip
\par

\section{Temperature-induced plasmon excitations in $1T^\prime$-MoS$_2$ monolayers}
\label{sec4}

 At finite temperatures, the dielectric function is determined by the temperature-dependent polarization function  $\Pi^{(0)}({\bf q},\omega \, \vert \, T)$. The latter quantity is obtained as an integral  transform of its zero-temperature counterpart, $\Pi^{(0)} (q,\omega \, \vert \, T=0)$ with the use of the Maldague formula \,\cite{SK30}

\begin{equation}
\Pi^{(0)}(q, \omega \vert T) = \frac{1}{2 k_B T} \, \int\limits_{-\infty}^{\infty}  d\eta \,
\frac{\Pi_{\phi,0} (q,\omega \, \vert \, \eta)}{1 + \cosh \left[(\mu - \eta) / k_B T \right] } \ ,
\label{Pi0T}
\end{equation}
where the integration is performed over all possible Fermi energies (doping levels) of the system at T=0 K. In order to be able to apply Eq.\   \eqref{Pi0T}, we need to know how the chemical potential $\mu(T)$ depends on temperature $T$.

\begin{figure}
\centering
\includegraphics[width=0.7\linewidth]{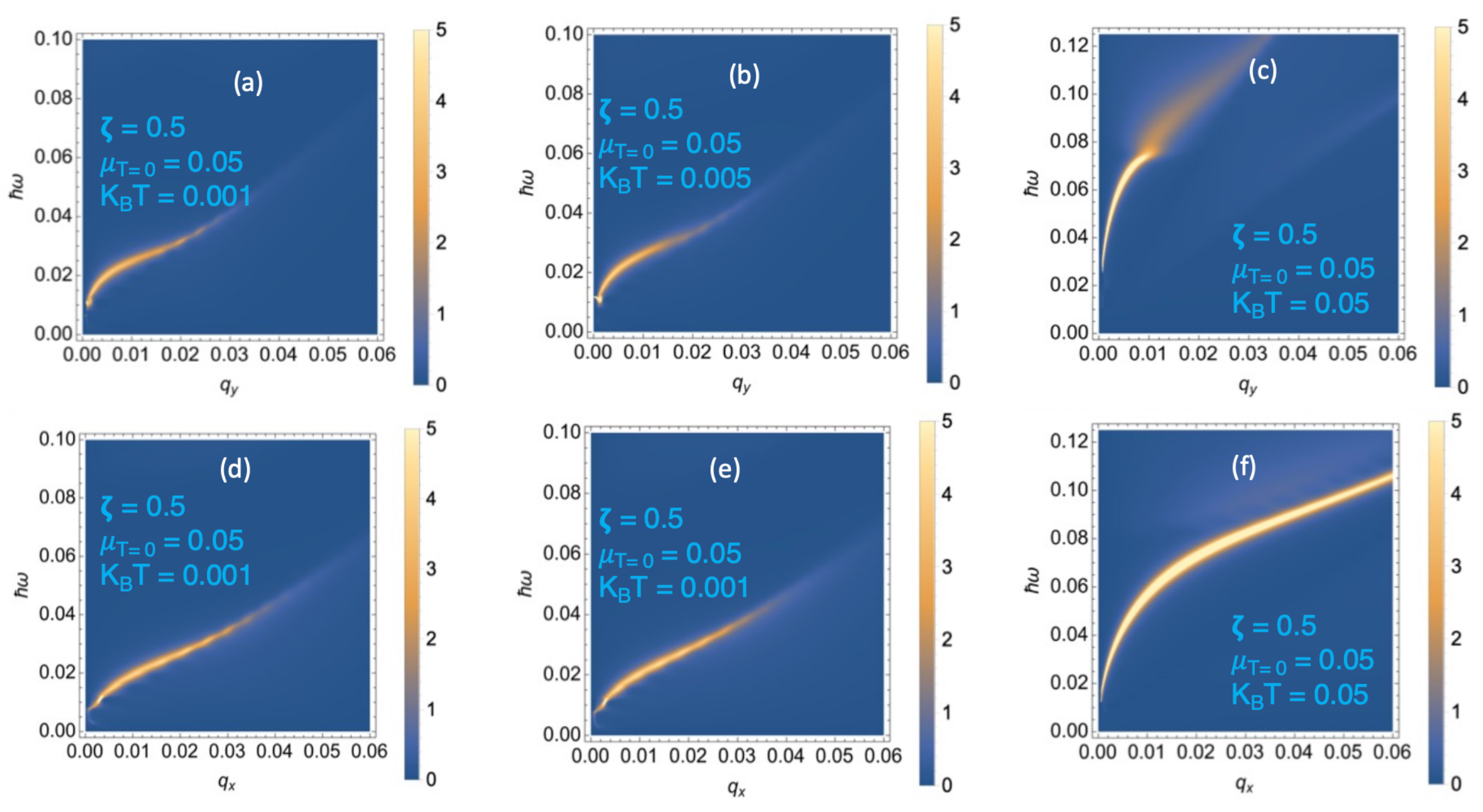}
\caption{(Color online) Density plots of plasmon dispersions obtained from the loss function $\text{Im}[1/\epsilon({\bf q}, \omega \vert T)]$ for $1T^\prime$-MoS$_2$ as a function of  the wave vector component $q_x$ and frequency $\omega$. The other component $q_y$ of the wave vector ${\bf q}$ is set equal to zero for all plots. Each panel corresponds to a different temperature as labeled. The peaks of the loss function correspond to a plasmon branch and the width of each peak is related to its damping. The Landau damping of the plasmon modes in the long wavelength limit is revealed.  The plasmons emerge from the particle-hole region at finite wave vector and frequency. }
\label{FIG:ai1}
\end{figure}

\begin{figure}
\centering
\includegraphics[width=0.7\linewidth]{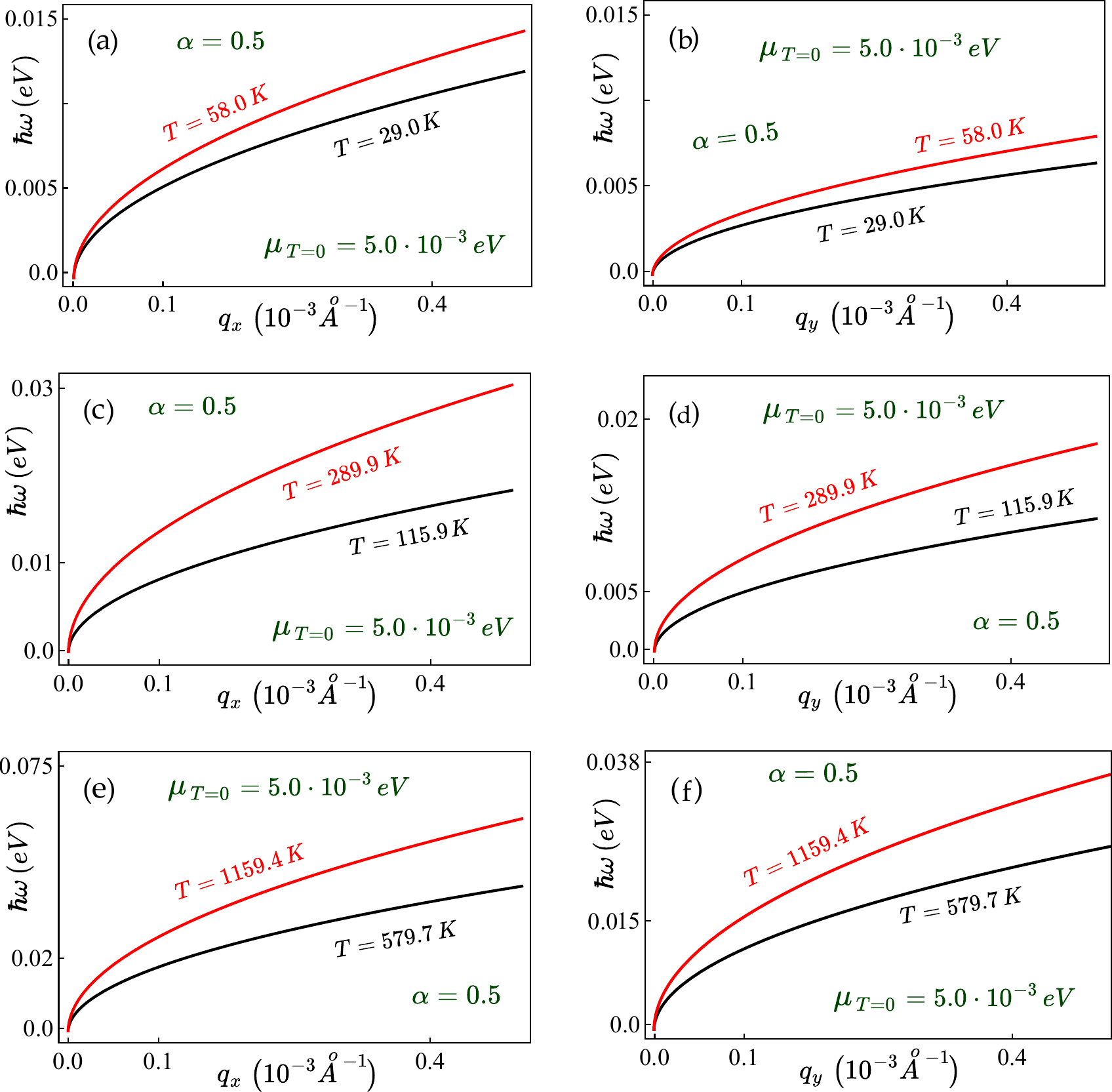}
\caption{(Color online) Calculated temperature-dependent plasmons in $1T^\prime$MoS$_2$. The frequency of each plasmon branch is presented as a function of the wave vector component $q_x$ for panels $(a)$, $(c)$ and $(e)$ on the left-hand side, and as functions of the other component $q_y$ for plots $(b)$, $(d)$ and $(f)$ on the right. In each panel, the two curves correspond to two chosen temperatures, as labeled. }
\label{FIG:ai2}
\end{figure}

\begin{figure}[!ht]
\centering
\includegraphics[width=0.8\linewidth]{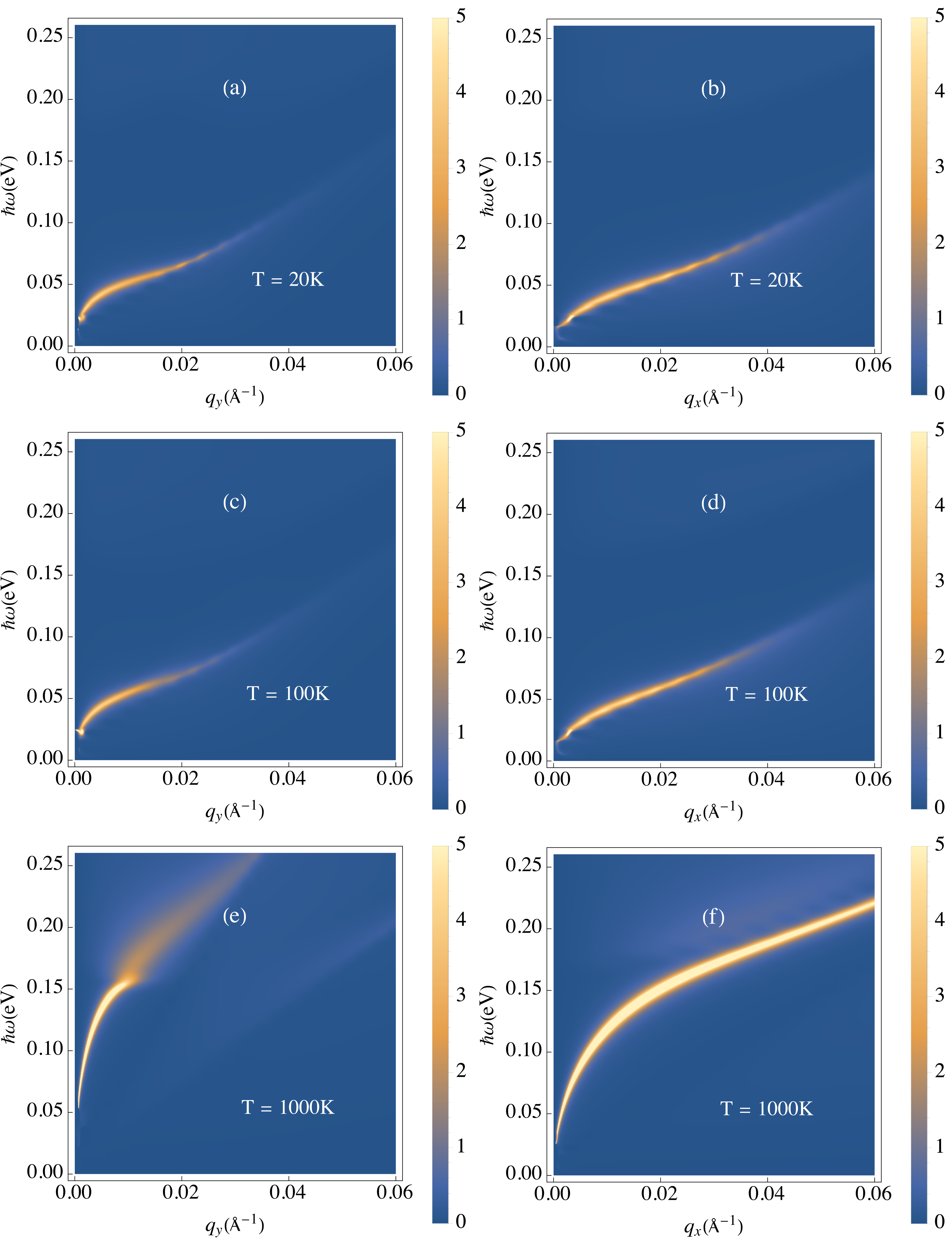}

\caption{ (Color online)  Plasmon dispersions at finite temperature for monolayer 1T$^\prime$MoS$_2$ in the presence of irradiation with irradiation parameter $\zeta=$0.2. The left panel [(a), (c) and (e)] shows plasmon dispersions obtained from the loss function along $q_y$ and the right panel  [(b), (d) and (f)]  along $q_x$ for different temperatures as labeled.  The chemical potential $\mu$ is chosen as 0.094eV and the vertical electric field is half of that critical field (i.e $\alpha=$0.5).}

\label{temp}
\end{figure}

\medskip
\par
Our numerical results for finite-temperature plasma  with no irradiation are presented in Figs.\ \ref{FIG:ai1} and \ref{FIG:ai2} through the loss function  $\text{Im}[1/\epsilon(q, \omega \vert T)]$. We note that the loss function depicts not only the locations of the plasmon branches, but also their decay rates $\gamma(\omega_p(q))$ via the width of each peak.

\medskip 
\par
 Our calculations have shown that  for chosen wave vector, the frequency of the plasmon in $1T^\prime$MoS$_2$  is increased as temperature is increased.  Qualitatively, we can say that a  $\backsimeq \sqrt{q \, T}$  law for graphene is also satisfied in the long wavelength limit for molybdenum disulfide. 

\medskip 
\par
Interestingly,  the decay rate due to Landau damping of the plasmon modes is reduced at higher temperatures, as we also see from Fig.~\ref{FIG:ai1}. We  can say that at the temperatures which are large compared to the Fermi energy,  the plasmon is once again well-defined over the larger range of wave vector $q$. The existence of a finite band gap in 1T$^\prime$MoS$_2$  also  results in the  existence of undamped (or slightly damped) plasmons  in the larger frequency range  compared to graphene. 

\medskip
\par
Furthermore, results for temperature-induced plasmon excitations with irradiation are shown in Fig.\ \ref{temp} for the same values of $\mu, \alpha$ and $\zeta$ which were chosen in Fig.\  \ref{loss_fn}. The plasmon dispersions in two different momentum directions are anisotropic which  resembles the anisotropic energy dispersion and the polarization along $q_y$ and $q_x$. We observe that the plasmons are with low frequency and weak intensity at low temperatures. For example, in Figs.\ \ref{temp}(a) and (b) for T$ = $20 K, finite loss function is seen in the  low momentum regime and  the plasmon dispersion tends to follow the $\sqrt{q}$ dispersion relation. However, the intensity  of the excitation spectrum is very low. As the temperature is increased from 20K, temperature induced excitations contribute to the polarization resulting in strong, stable and undamped plasmons with  $\sqrt{q}$ dispersion  in the long wavelength region as shown in Figs.\   \ref{temp}(c) and (d) for T$=$ 100 K.  The limit of the wavelength to which plasmons exist is different along the two different momentum directions. At high temperatures, the plasmon frequency is high and the lifetimes are also increased along $q_{x}$ but the lifetimes of the plasmons are very short along $q_{y}$ which can be seen in Figs.\  \ref{temp}(e) and (f) for T$=$  1000 K.

\medskip 
\par
We have also observed that anisotropy and tilting of the energy bands of 1T$^\prime$MoS$_2$ as having a significant effect on the plasmon dispersion relation at all temperatures. The frequencies of the plasmon modes, for chosen values of $q_x$ and $q_y$, differ by almost 100\%, and this effect is still present at the higher temperatures as well.

\section{Application of the Polarization Function to the Exchange Energy}
\label{sec5}

The kinetic, exchange and correlation energies  are key ingredients in the calculations of several physical properties. These include  the   ground-state energy and compressibility of Dirac electrons in 1T$^\prime$-MoS$_2$ within the random phase approximation at zero temperature. The calculation of the compressibility of two dimensional (and three dimensional) materials has been the object of recent calculations on materials with  Dirac cone band structure.\cite{SSC}  In the non-interaction limit, the compressibility of a Fermi liquid can be expressed in terms of quantities which are related to its band structure. Furthermore, the compressibility may be experimentally probed since it is related to the quantum capacitance of the electron liquid which can be determined with the use of a scanning tunneling microscope. Another interesting application of the exchange energy has been in the calculation of the scattering rates of the Coulomb excitations.  In Ref. [\onlinecite{GG2}], the de-excitation processes were studied using the screened exchange energy.  That calculation employed the intraband single-particle excitations (SPEs), the interband SPEs, and the plasmon modes, depending on the quasiparticle states and the Fermi energy. These works motivated us to calculate the exchange energy of 1T$^\prime$MoS$_2$. However the calculation of the correlation energy demands an enormous amount of computing  which is ongoing.

\medskip
\par

From many-body theory\, \cite{GG1}, we know that the total Coulomb interaction energy for electrons in a two-dimensional material can be expressed as\,\cite{SK36}

\begin{equation}
\mathbb{E}_{int} = \frac{N}{2} \int\limits_{0}^{1}\,d \chi\, \int \frac{d^2 {\bf q}}{(2 \pi)^2} V(q) \left[
\mathbb{S}(\chi, {\bf q})-1 \right] \ ,
\label{Eint}
\end{equation}
where the Coulomb energy for the uniform positive charged background is deducted, $N$ is the total number of electrons in the system.  The dynamical structure factor $\mathbb{S}(\chi, {\bf q})$ introduced in Eq.\,  (\ref{Eint}) can be obtained from the electron polarization function of \textit{imaginary frequency} calculated within the random-phase approximation. This yields

\begin{equation}
\mathbb{S}(\chi, {\bf q}) = - \frac{\hbar}{\pi n} \int\limits_{0}^{\infty} d\omega \, \Pi^{\chi}({\bf q},i\omega) \ ,
\end{equation}
where $n$ is the electron density and the screened polarization function  is given by

\begin{equation}
\Pi^{\chi}({\bf q},i\omega ) = \frac{\Pi^{(0)}({\bf q},i\omega)}{1-\chi V(q) \Pi^{(0)}({\bf q},i\omega)}\ .
\label{eqn-24}
\end{equation}
 
\begin{figure}[ht]
\centering
\includegraphics[width=0.8\linewidth]{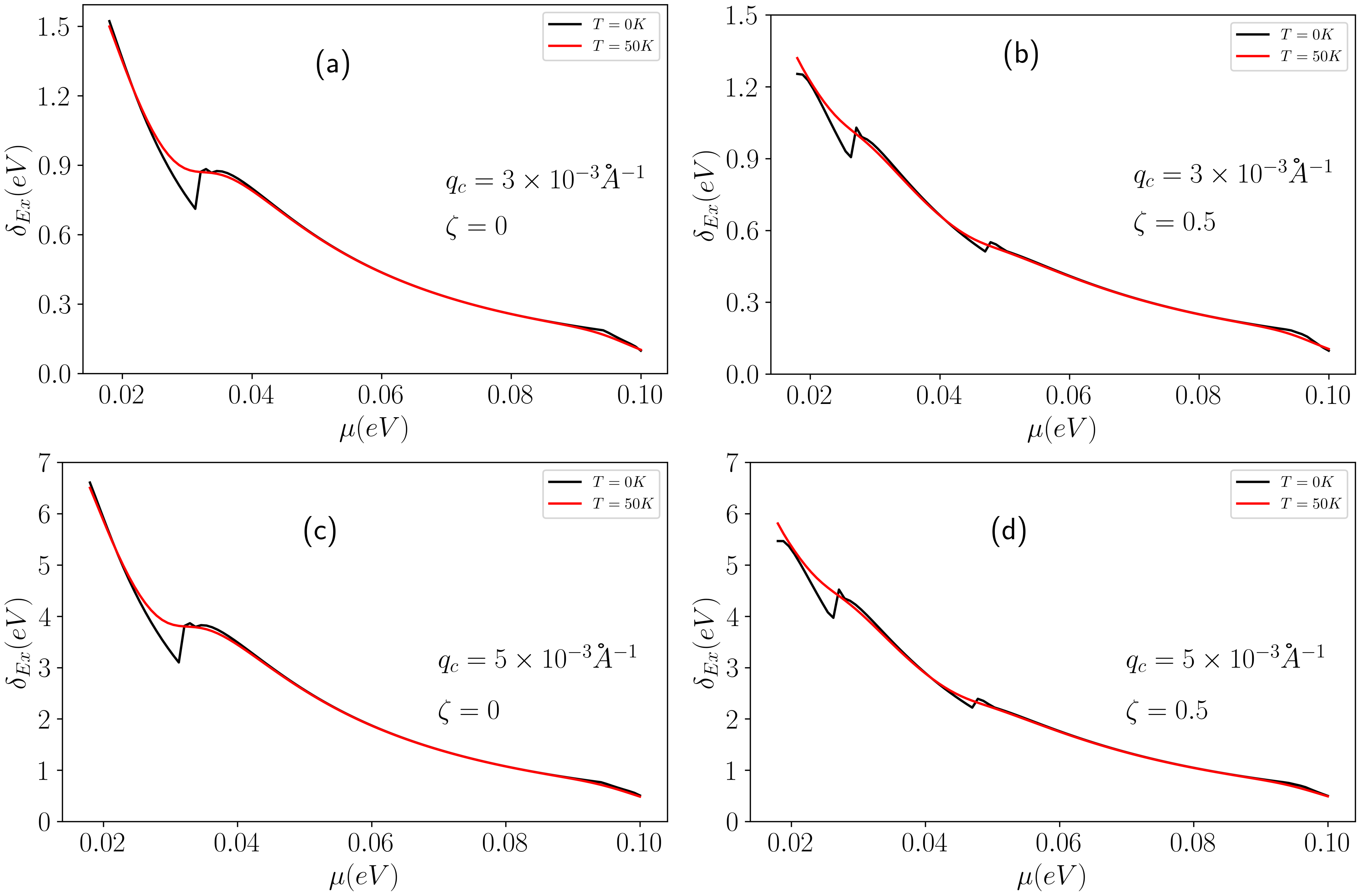}
\caption{(Color online) The exchange energy $\delta E_x $  as a function of  chemical potential $\mu$ for chosen perpendicular electric field parameter $\alpha = 0.5$ and temperature $T$ as indicated.  The graphs in the left panel are for undressed states and those in the right panel are for dressed states with $\zeta =0.5$. Plots (a) and (b) are for cutoff  values $q_{c} = 3\times 10^{-3} \AA^{-1}$. Panels (c) and (d) are for  $q_{c} = 5\times 10^{-3} \AA^{-1}$.  The $\mu$ has been changed with doping, keeping the temperature constant.  }
\label{ex_a}
\end{figure}

\medskip
\par

The interaction energy in Eq.\,  (\ref{Eint}) appears to be divergent, which is attributed to the fact that the calculation includes an interaction with an infinite sea of holes.\,\cite{EXC1,Asgari} To remove this divergence, we choose the total energy of undoped 1T$^\prime$MoS$_2$  ($E_F=0$ in all channels) as our point of reference. Incorporating this reference point into our calculations, we will be  able to separate the total interaction energy per electron into exchange $\mathbb{E}_x$ (for $  \chi= 0$) and correlation $\mathbb{E}_c$ contributions. We now separate the contributions from the exchange and correlation energies and write

\begin{equation}
\mathbb{E}_x   =  - \frac{\hbar}{2 \pi n} \int  d{\bf r} \int d{\bf r}^\prime  \  v({\bf r}-{\bf r} ^\prime ) \left[ \int  \frac{d\omega}{2\pi} \Pi^{(0)} \left( {\bf r}-{\bf r} ^\prime   ; i\omega    \right) +n({\bf r}) \delta({\bf r}-{\bf r} ^\prime )
\right]
\label{ex1}
\end{equation}

\begin{eqnarray}
\mathbb{E}_c   &=& \mathbb{E}_{xc}  - \mathbb{E}_x
\nonumber\\
&=&   \int d{\bf r}  \int_0^1   d\Lambda \left[  n({\bf r})  -  n_\Lambda({\bf r}) \right] 
\left[  \upsilon_{ext}({\bf r})  -\phi({\bf r})
\right] 
\nonumber\\
&- &  \frac{1}{2}  \int d{\bf r} \int d{\bf r}^\prime  v({\bf r}-  {\bf r}^\prime) \int_0^1  d\Lambda
\int_0^\infty  \frac{d\omega}{\pi}  \tilde{\Pi}^\Lambda  ({\bf r} ,{\bf r}^\prime ;i\omega )  \    ,  
\label{Exc1}
\end{eqnarray}
where

\begin{equation}
\tilde{\Pi}^\Lambda ({\bf r},{\bf r}^\prime ;i\omega) \equiv
\Pi^\Lambda({\bf r},{\bf r}^\prime ;i\omega) -\Pi^{(0)}({\bf r},{\bf r}^\prime ;i\omega)
\end{equation}

\begin{equation}
\phi_\Lambda({\bf r})  \equiv  \frac{1}{2}   \int d{\bf r}^\prime
v({\bf r}-{\bf r}^\prime) \left[ n({\bf r}^\prime) +n_\Lambda({\bf r}^\prime)
-2N({\bf r}^\prime) \right]
\end{equation}
Then, taking into account the total energy of undoped 1T$^\prime$MoS$_2$  ($E_F=0$ in all channels) as  our reference point energy and introducing Fourier transforms, we have the excess exchange energy and excess correlation energy per excess electron of doped 1T$^\prime$MoS$_2$ 

\begin{equation}
\delta \mathbb{E}^{RPA}_x   =  - \frac{\hbar}{2 \pi n} \int \frac{d^2 {\bf q}}{(2 \pi)^2 } V(q) \int\limits_{0}^{\infty} d\omega \, \delta\Pi^{(0)}({\bf q} ,i\omega; \mu) 
\ , 
\label{eqn-28} 
\end{equation}
where we introduce the following notation

\begin{equation}
\delta\Pi^{(0)}({\bf q} ,i\omega; \mu) =  \Pi^{(0)}({\bf q},i\omega; \mu) - \Pi^{(0)}({\bf q} ,i\omega; \mu \le G(\lambda,\alpha, \xi)) \ ,
\label{eqn-28b}
\end{equation}
and

\begin{equation}
\delta \mathbb{E}_c^{RPA} =  \frac{\hbar}{2 \pi n} \int \frac{d^2 {\bf q}}{(2 \pi)^2 } \int\limits_{0}^{\infty} d\omega \,
 \left\{ V(q) \, \delta\Pi^{(0)}({\bf q} ,i\omega;\mu) + \ln[\mathbb{A}({\bf q} ,i\omega; \mu)] \right\} \, ,
\label{eqn-29}
\end{equation}
where

\begin{equation}
\mathbb{A}({\bf q},i\omega; \mu) = \frac{1-V(q) \, \Pi^{(0)}({\bf q},i \omega; \mu)}{1-V(q) \, \Pi^{(0)}({\bf q}, i \omega; \mu \le G(\lambda,\alpha, \xi)}\ .
\end{equation}
In the calculation of exchange energy we have a frequency integral to do  which can be carried out in a straightforward way by making use of 

\begin{equation}
\mbox{\texttt{Re}}\int_0^\infty  d\omega 
\frac{1}{\epsilon_{k+q}-\epsilon_q -i\omega}=-\frac{\pi}{2\hbar}
\mbox{\texttt{sgn}} (\Omega_{kq})\  ,
\end{equation}
where $\hbar\Omega_{kq}\equiv \epsilon_k-\epsilon_{k+q}$.

\medskip
\par

In this paper, we are interested in calculating numerically the exchange energy of tilted 1T$^{\prime}$MoS$_2$. Clearly,  the effects from both exchange and correlation tend to cancel each other if the small logarithmic correction can be neglected. As seen above in Eq.\ (\ref{eqn-28}) and Eq.\ (\ref{eqn-29}), $\mbox{\texttt{Re}}\,\Pi^{(0)}(q, i \omega;\mu)$ with imaginary frequency  plays an important role in the calculation of the many-body effect on the total energy of the system since its imaginary part is zero for real frequencies. Furthermore, from Eq.\ (\ref{eqn-28b}), it is seen that the polarization for $\mu$ less than or equal to the gap $G(\lambda,\alpha, \xi)$ is subtracted out. This means that the exchange energy does not depend on  the first term of the polarization in Eq.\ (\ref{pseparate}) and hence only the excess carriers due to doping contribute to the final result. There have been a number of reported works dealing with the derivation of these equations for exchange and correlation  in the RPA.  The presentation of analytical expressions for the noninteracting polarization function $\Pi^{(0)}({\bf q},\omega)$ with  real frequency\,\cite{SK45,SK36,SK47,SK38,SK19} has also been made. All of them used a delta function representation for the imaginary part of $\Pi^{(0)}({\bf q},\omega)$, which is a consequence of an infinitesimally  small broadening added to the real frequency  $\omega \Longrightarrow \omega +i\delta^+$. The first angular integral with respect to ${\bf k}$ can be easily performed at zero temperature, which results in a collection of Heaviside step functions for the single-particle excitation regions. However, this approach is not applicable in the case of finite imaginary frequency, $\omega \Longrightarrow i\omega$. In this case, we analytically carried out the integration over $\omega$ and a numerical method is employed for the integration over momentum space which is free from any  approximations.  

\medskip
\par

The numerical results presented in Fig.\  \ref{ex_a} illustrate the dependence of the exchange energy on doping associated with $\mu$, temperature and  wave vector cutoff $q_{c}$ for chosen vertical electric field $\alpha$ and spin-orbit coupling gap $\Delta$. The exchange energy is positive regardless of either the positive or negative energies due to doping. We observe that the exchange energy is decreased with increasing $\mu$ and possesses a sharp kink at zero temperature as the Fermi level crosses a subband.  This is the main difference between the exchange energy in other 2D isotropic materials (graphene \cite{EXC1}) and anisotropic tilted band 1T$^{\prime}$MoS$_2$. As the density of states sharply falls in the spin separated gap region, the interaction energy tends to increase. Consequently, a sharp kink appears at absolute zero. However at finite temperature, thermally excited electrons enhance the density of states in this region and hence the interaction energy varies smoothly. with the chemical potential. While considering temperature effects, it is important to mention that we have adjusted the temperature keeping the electron population conserved. The exchange energy per excess electron vanishes for heavy doping. In contrast, in the presence of irradiation, the two kinks in plots (b) and (d) of Fig. \  \ref{ex_a}, small in comparison with that in the absence of irradiation, clearly reflects the features of  the energy band and the reshuffling spin band gap. Since for the dressed states the spins are separated with unequal energy in two different valley, the values are not equal when adding up  their contributions from both valley as it is for the undressed states. Therefore, these two kinks  correspond to the two different spin bandgaps in two opposite Dirac points. We did not observe significant effects due to temperature in the exchange energy  in any other region except in this sharply rise region. The figure further explains that the exchange energy linearly increased with wave vector cutoff $q_{c}$ until the saturation point is achieved which is also true for the 2D electron gas and graphene \cite{EXC1}. This verifies that we do not need to integrate over the entire $q$-space.

\medskip
\par

\medskip
\par
From the preceding discussion, we conclude that the exchange energy for 1T$^{\prime}$MoS$_2$  can be tuned. This may be accomplished with electron density, temperature and irradiation and a careful selection of vertical electric field, momentum cutoff and spin-orbit coupling gap. Sufficient information on the exchange and correlation energy is required for the material fabrication as it affects different electronic properties such as the Hall conductance and magnetic susceptibility. Higher order values of exchange and correlation energies are required for the material to be useful for technological purposes.\cite{EXC1}

\medskip
\par

\section{Concluding remarks}
\label{sec6}

In this paper, we have meticulously established a rigorous theoretical program for Floquet engineering, investigating and subsequently tailoring the electronic properties of monolayer semiconducting  1T$^\prime$MoS$_2$.  An external high-frequency dressing field was employed within the off-resonance regime. We have demonstrated that monolayer 1T$^\prime$MoS$_2$ displays tunable and gapped spin- and valley-polarized tilted Dirac bands, thereby showing the full complexity of the low-energy Hamiltonian for  the non-irradiated material. We have calculated and analyzed the properties of the electron dressed states corresponding to circular polarization of a dressing field by focusing on their symmetry, anisotropy, tilting, direct and indirect band gaps.   With the use of these results, we have calculated numerically the dynamic polarization function, the plasmon excitation spectrum as well as the exchange energy for tilted 1T$^\prime$MoS$_2$ with and without circularly polarized irradiation. We observe that the energy bands are anisotropic in two perpendicular directions. The tilting of the bands is observed in one. Direction. The polarization and plasmons are also anisotropic accordingly.

\medskip
\par
Our calculations revealed that plasmons supported by 1$T^{\prime}$Mo$S_{2}$ are not the only plasmonic phenomena in the realm of 2D materials shown to be impacted by the anisotropy of the energy band structure\cite{silkin}. The dynamical dielectric response strongly depends on the direction of the in-plane wave vector ${\bf q}$. Additionally, in contrast to conventional 2DEG with  a plasmon dispersion $\sim \sqrt{q}$ in the long wavelength limit, 1$T^{\prime}$Mo$S_{2}$ has a gap in its dispersion. This offers unique characteristics and potential applications. The anisotropy property can be exploited for directional control of plasmon propagation, leading to novel functionalities in nanophotonic devices. Next, the effect of irradiation on the polarization and plasmon dispersion of 1$T^{\prime}$Mo$S_{2}$ has also been explored. It has not been studied even in the $8-P mmn $ borophene, which also has tilted anisotropic energy dispersion. In this paper, it has been demonstrated  that the polarization and plasmon frequency are suppressed in the presence of a circularly polarized electromagnetic field. However, the plasmons are sustained over a  wider range than in the absence of irradiation.  Additionally, comparison between non-irradiated and circularly polarized light irradiated 1T$^\prime$MoS$_2$ allows us to understand that the dressing field can tune the bandgap, suppress the polarization and weaken the intensity of the plasmon excitation.  Also, it reduces the exchange energy in monolayer tilted-band $1T^\prime$MoS$_2$.  The temperature-induced polarization  and the finite-temperature behavior of plasmon excitations in monolayer 1T$^\prime$MoS$_2$ were also studied.  Although we have not explicitly studied the optical  properties of 1$T^{\prime}$Mo$S_{2}$, we believe that its optical conductivity will reveal interesting behaviors  due in  part to its relationship with the polarization function, i.e., \(\Pi^{(0)}({\bf  q},\omega) \sim \sum_{i, j}\sigma_{ij}(\omega)q_iq_j\). The optical properties can be tuned with different parameters like the vertical electric field, irradiation strength and temperature.  In addition, studying plasmons in 1$T^{\prime}$Mo$S_{2}$ complements research on plasmons in other 2D materials like graphene, the transition metal dichalcogenides (TMDs), and black phosphorus. Understanding the similarities and differences between plasmons in these materials helps in gaining a broader knowledge of plasmonic phenomena.

\medskip
\par

\medskip
\par
It has been shown that the exchange energies depend on the polarization with imaginary frequency and reflect the essential features of the energy band dispersions.  So far, we have not been able to determine the effect due to exchange on the plasmon excitations, which is a higher order effect.  The plasmon dispersion relation was calculated using the mean-field RPA.  Going beyond the RPA, one will have to include exchange and correlation effects  through Green’s functions.  This was beyond the scope of our calculations.

\begin{acknowledgements}
G.G. would like to acknowledge Grant No. FA9453-21-1-0046 from the Air Force Research Laboratory (AFRL). 
\end{acknowledgements}


\clearpage


\begin{thebibliography}{69}


\bibitem{AI1}  C. J. Tabert and E. J. Nicol, Valley-Spin Polarization in the Magneto-Optical Response of Silicene and Other Similar 2D Crystals,  Phys.  Rev.  Lett.  {\bf 110}, 197402 (2013).

\bibitem{AI2}   A. Iurov, L. Zhemchuzhna, P. Fekete, G. Gumbs, and D. Huang, Klein tunneling of optically tunable Dirac particles with elliptical dispersions, Phys.   Rev.  Research  {\bf 2}, 043245 (2020).

\bibitem{AI3}   M. Islam and S. Basu,  Spin and charge persistent currents in a Kane Mele $\alpha$-T$_3$ quantum ring,   Journal of Physics: Condensed Matter, {\bf 36}, 135301  (2023).  

\bibitem{AI4}   A. Iurov, G. Gumbs, and D. Huang, Peculiar electronic states, symmetries, and Berry phases in irradiated  $\alpha$-T$_3$,    Phys.   Rev.  B {\bf  99}, 205135 (2019). 


\bibitem{AI5}  M. Islam, T. Biswas, and S.  Basu, Effect of magnetic field on the electronic properties of an $\alpha$-T$_3$ ring,  Phys.  Rev.  B {\bf 108}, 085423 (2023).


\bibitem{AI6}   E. Gorbar, V. Gusynin, and D. Oriekhov,   Electron states for gapped pseudospin-1 fermions in the field of a charged impurity,  Phys . Rev.  B {\bf 99}, 155124 (2019).

\bibitem{AI7}   E. Illes and E. Nicol,  Klein tunneling in the  $\alpha$-T$_3$ model, Phys.  Rev.   B {\bf  95},  235432 (2017).

\bibitem{AI8}   C.-X. Yan, C.-Y. Tan, H. Guo, H.-R. Chang, et al., Highly anisotropic optical conductivities in two-dimensional tilted semi-Dirac bands, Phys.  Rev.  B {\bf 10}, 195427  (2023).

\bibitem{AI9}   A. Iurov, L. Zhemchuzhna, D. Dahal, G. Gumbs, and D. Huang, Quantum-statistical theory for laser-tuned transport and optical conductivities of dressed electrons in $\alpha$-T$_3$ materials,  Phys.  Rev.  B  {\bf 101}, 035129 (2020)

\bibitem{AI10}   C.-Y. Tan, C.-X. Yan, Y.-H. Zhao, H. Guo, H.-R. Chang, et al.,  Anisotropic longitudinal optical conductivities of tilted Dirac bands in 1T$^\prime$- MoS$_2$, Phys.  Rev.   B  {\bf 103}, 125425 (2021).

\bibitem{AI11}   Y. Gomes and R. O. Ramos,  Tilted Dirac cone effects and chiral symmetry breaking in a planar four-fermion model, Phys.  Rev.  B  {\bf 104}, 245111 (2021).


\bibitem{AI12}    J.  Carbotte, K. Bryenton, and E. Nicol,   Optical properties of a semi-Dirac material, Phys.  Rev.  B {\bf  99}, 115406 (2019).


\bibitem{AI13}  S. K. Firoz Islam and Arijit Saha,   Driven conductance of an irradiated semi-Dirac material Phys.  Rev.  B {\bf  98}, 235424 (2018). 

 \bibitem{AI14}   Q.-Y. Xiong, J.-Y. Ba, H.-J. Duan, M.-X. Deng, Y.-M. Wang, and R.-Q. Wang,  Optical conductivity and polarization rotation of type-II semi-Dirac materials, Phys.  Rev.  B  {\bf 107}, 155150  (2023).


\bibitem{AI15}   P.-H. Shih, G. Gumbs, D. Huang, A. Iurov, and Y. Abranyos,  Blocked electron transmission/reflection by coupled Rashba–Zeeman effects for forward and backward spin filtering, Journal of Applied Physics {\bf 132},  154302  (2022).

\bibitem{AI16}   X.-F. Wang,  Plasmon spectrum of two-dimensional electron systems with Rashba spin-orbit interaction, Phys.  Rev.  B {\bf  72}, 085317   (2005).

\bibitem{AI17}    K. Kristinsson, O. V. Kibis, S. Morina, and I. A. Shelykh, Control of electronic transport in graphene by electromagnetic dressing,  Scientific  Reports  {\bf  6}, 1 (2016).

\bibitem{AI18}   O. Kibis,   Metal-insulator transition in graphene induced by circularly polarized photons,   Phys.  Rev. B {\bf  81}, 165433 (2010).

\bibitem{ANA1}  W. Choi, N. Choudhary, G.H. Han, J. Park, D. Akinwande and Y.H. Lee, Recent development of two-dimensional transition metal dichalcogenides and their applications,  Materials Today {\bf 20}  number 3  (2017).

\bibitem{ANA4}   G.  Grosso, S. B. Chand, J. M. Woods, and E. Mejia, Controlling Dark Excitons in Two-Dimensional TMDs for Optoelectronic Applications, in 2D Photonic Materials and Devices VI, edited by A. Majumdar, C. M. T. Jr, and H. Deng, Vol. PC12423 (SPIE, 2023).

\bibitem{ANA5}    S. B. Chand, J. M. Woods, J. Quan, E. Mejia, T. Taniguchi, K. Watanabe, A. Al`u, and G. Grosso,  Interaction-driven transport of dark excitons in 2D semiconductors with phonon-mediated optical readout, Nature Commun.  {\bf 14}, 3712 (2023).

\bibitem{ANA6}   S. B. Chand, J. M. Woods, E. Mejia, T. Taniguchi, K. Watanabe, and G. Grosso, Visualization of Dark Excitons in Semiconductor Monolayers for High-Sensitivity Strain Sensing, Nano Lett. {\bf 22}, 3087 (2022).
	
\bibitem{ANA7}    Y. M. P. Gomes and R. O. Ramos, Superconducting phase transition in planar fermionic models with Dirac cone tilting,  Phys. Rev. B {\bf107}, 125120 (2023).

\bibitem{ANA8}     A. Balassis, G. Gumbs, and O. Roslyak, Polarizability, plasmons, and screening in 1T$^\prime$-MoS$_2$ with tilted Dirac bands, Phys. Lett.  A {\bf 449}, 128353,  (2022). 

\bibitem{ANA410}     K.  F.  Mak  and  J. Shan,  Photonics and optoelectronics of 2D semiconductor transition metal dichalcogenides, Nat. Photonics. {\bf 10}, 216 (2016).

\bibitem{Guinea}  A. H. Castro Neto, F. Guinea, N. M. R. Peres, K. S. Novoselov, and A. K. Geim,  The electronic properties of graphene,  Rev. Mod. Phys. {\bf 81}, 109 (2009).

\bibitem{Ando}  T. Ando,  A. B. Fowler, and F. Stern, Electronic properties of two-dimensional systems, Rev. Mod. Phys. {\bf 54}, 437 (1982).

\bibitem{IOP} G. G. Naumis, S. Barraza-Lopez, M. Oliva-Leyva, and H. Terrones,   Electronic and optical properties of strained graphene and other strained 2D materials: a review,  Reports on Progress in Physics  {\bf 80}, number 9 (2017)     
 
\bibitem{Gulley} J. R. Gulley and D. Huang, Self-consistent quantum-kinetic theory  for interplay between pulsed-laser excitation and nonlinear  carrier transport in a quantum-wire array, Opt. Express OE {\bf 27}, 17154 (2019).

\bibitem{plas1} O. L. Berman, G. Gumbs, and Y. E. Lozovik,  Magnetoplasmons in layered graphene structures,  Phys. Rev. B {\bf 78}, 085401  (2008).

\bibitem{plas2}  H. Yan, Z. Li, X. Li, W. Zhu, P. Avouris, and F. Xia, Infrared Spectroscopy of Tunable Dirac Terahertz Magneto-Plasmons in Graphene,  Nano letters  {\bf 12}, 3766 (2012).

\bibitem{plas3} J.-Y. Wu, S.-C. Chen, O. Roslyak, G. Gumbs, M.-Fa Lin,  Plasma excitations in graphene: Their spectral intensity and temperature dependence in magnetic field,  ACS nano {\bf  5},  1026 (2011).  

\bibitem{plas4} R. Roldan, J.-N. Fuchs, and M. Goerbig,   Collective modes of doped graphene and a standard two-dimensional electron gas in a strong magnetic field: Linear magnetoplasmons versus magnetoexcitons,  Phys.  Rev. B {\bf 80}, 085408 (2009).

\bibitem{Nafis}  A. N.  Arafat, O. L. Berman, and G. Gumbs,    Superfluidity of indirect momentum space dark excitons in a double layer of 1T$^\prime$-MoS$_2$ with tilted semi-Dirac bands, Phys. Rev. B  {\bf  109},  2245  (2024).     

\bibitem{Berry} Li-Li Ye, C.-Z.  Wang, and Y.-C. Lai,  Experimental scheme for determining the Berry phase in two-dimensional quantum materials with a flat band, Phys. Rev. B {\bf 110}, 075108   (2024).

\bibitem{EXC1} Y. Barlas, T. Pereg-Barnea, M. Polini, R. Asgari, and A. H. MacDonald,  Chirality and Correlations in Graphene,  Phys. Rev. Lett. {\bf 98}, 236601,  (2007).

 \bibitem{silkin} U. Muniain  and V. M. Silkin ,  Impact of the energy dispersion anisotropy on the plasmonic structure in a two-dimensional electron system,  Phys. Chem. Chem. Phys. {\bf  24},17885 (2022).

\bibitem{SK38} P. K. Pyatkovskiy, Dynamical polarization, screening, and plasmons in gapped graphene,  J. Phys.: Condens. Matt. {\bf 21}, 025506 (2009).

\bibitem{SK19} S. Das Sarma and Q. Li,  Intrinsic plasmons in two-dimensional Dirac materials, Phys. Rev. B {\bf 87}  235418 (2013).

\bibitem{SK20} V. -M. Nguyen, Temperature and inhomogeneity combination effects on collective excitations in three-layer graphene structures, Physica E: Low-dimensional Systems and Nanostructures {\bf 140}, 115201 (2022).

\bibitem{SK21}A. Iurov, G. Gumbs, D. Huang, and V. M. Silkin, Plasmon dissipation in gapped graphene open systems at finite temperature, Phys. Rev. B {\bf 93}, 035404 (2016).

\bibitem{SK22} D. K. Patel,  S. S. Ashraf, and  A. C. Sharma, Finite temperature dynamical polarization and plasmons in gapped graphene, Physica Status Solidi (b) {\bf 252 (8)}, 1817-1826 (2015).

\bibitem{SK23} D. V. Tuan, and N. Q. Khanh, Plasmon modes of double-layer graphene at finite temperature, Physica E: Low-dimensional Systems and Nanostructures {\bf 54}, 267-272 (2013).

\bibitem{SK24} A. Iurov, G. Gumbs, D. Huang, and G.  Balakrishnan, Thermal plasmons controlled by different thermal-convolution paths in tunable extrinsic Dirac structures, Phys. Rev. B {\bf 96}, 245403 (2017).

\bibitem{SK25} S. M. Badalyan, and F. M. Peeters,  Effect of nonhomogenous dielectric background on the plasmon modes in graphene double-layer structures at finite temperatures, Phys. Rev. B {\bf 85}, 195444 (2012).

\bibitem{SK26} A. Iurov, L. Zhemchuzhna, G. Gumbs, D. Huang, D. Dahal, and Y. Abranyos  Finite-temperature plasmons, damping, and collective behavior in the $\alpha-T_3$ model, Phys. Rev. B {\bf 105}, 245414 (2022).

 \bibitem{PRB9} H. Ibach and D.L. Mills, Electron Energy Loss Spectroscopy and Surface Vibrations, Academic Press: Cambridge, MA, USA, 2013.
 
 \bibitem{PRB10} R. Brydson, Electron Energy Loss Spectroscopy, Garland Science, New York, USA, 2020.

 \bibitem{PRB11} J. Lu and K. P. Loh, Plasmon dispersion on epitaxial graphene studied using high-resolution electron energy-loss spectroscopy, Phys. Rev. B {\bf 80} 113420 (2009). 

\bibitem{SK30} P. F. Maldague,  Many-body corrections to the polarizability of the two-dimensional electron gas, Surface Science {\bf 73}, 296 (1978).

 \bibitem{SK31} N. M. R. Peres, F. Guinea, and A. H. Castro Neto, Coulomb interactions and ferromagnetism in pure and doped graphene, Phys. Rev. B  {\bf 72},  174406 (2005).

\bibitem{SK32} J. Martin,  N. Akerman, G. Ulbricht,  T. Lohmann,    J. H. Smet,    K.von Klitzing, A. Yacoby,  Observation of electron–hole puddles in graphene using a scanning single-electron transistor, Nat. Phys. {\bf 4}, 144 (2008).

\bibitem{PRB 5} Q. Cheng, Y. pan, H. Wang, C. Zhang, D. Yu, A. Gover, H. Zhang, T. Li, L. Zhou, and S. Zhu, Observation of Anomalous $pi$ Modes in Photonic Floquet Engineering, Phys. Rev. Lett. {\bf 122} 173901 (2019).

\bibitem{PRB6} J. W. Mclver, B. Schulte, F.-U. Stein, T. Matsuyama, G. Jotzu, G. Meier and A. Cavalleri, Light-induced anomalous Hall effect in graphene, Nature Phys. Lett. {\bf 16} 38-41 (2020).

\bibitem{PRB6a}  J. Karch, P. Olbrich, M. Schmalzbauer, C. Zoth, C. Brinsteiner et. al, Dynamic Hall Effect Driven by Circularly Polarized Light in a Graphene Layer, Phys. Rev. Lett. {\bf105} 227402 (2010).

\bibitem{PRB7} G. Jotzu, M. Messer, R. Desbuquois, M. Lebart, T. Uehlinger, D. Greif and T. Esslinger, Experimental realization of the topological Haldane model with ultracold fermions, Nature {\bf 515} 237-240 (2014).

\bibitem{PRB8} C. Weitenberg and J. Simonet, Tailoring quantum gases by Floquet engineering, Nature Phys. {\bf 17} 1342-1348 (2021).

\bibitem{GG2} P.-H. Shih, C.-W. Chiu, J.-Y. Wu, T.-N. Do, and M.-Fa Lin, Coulomb scattering rates of excited states in monolayer electron-doped germanene, Phys. Rev. B {\bf 97}, 195302 (2018).

\bibitem{SK34} Z. Jalali-Mola, S. Jafari, Polarization tensor for tilted Dirac fermion materials : covariance and deformed Minkowski space- time, Phys. Rev. B {\bf 100}  075113  (2019).

\bibitem{Kalantar} Q. H. Wang, K. Kalantar-Zadeh, A. Kis, J. N. Coleman, and M. S. Strano, Highly anisotropic and robust electronic properties of monolayer $MoS_2$ due to strained sulfur-metal coordination, Nano Lett. {\bf 12} 9 (2012).

\bibitem{SK18b} M. Busl, G. Platero, and A. P. Jauho, Dynamical polarizability of graphene irradiated by circularly polarized ac electric fields, Phys. Rev. B {\bf 85 }, 155449 (2012).

\bibitem{GG1} G. Gumbs and D. Huang, Properties of Interacting Low-Dimensional Systems, John Wiley and Sons, (2013).

\bibitem{fengping} S. Yuan, F. Jin, R. Roldan, A.-P. Jauho and M. I. Katsnelson, Screening and collective modes in disordered graphene antidot lattices, Phys. Rev. B {\bf 88}, 195401 (2013).

\bibitem{malcolm} J. Malcolm, E. Nicol, Frequency-dependent polarizability, plasmons, and screening in the two-dimensional pseudospin-1 dice lattice, Phys. Rev. B {\bf 93}, 165433 (2016).

 \bibitem{SSC} D. S. L. Abergel,  Compressibility of graphene, Solid State Commun, {\bf  152},  1383  (2012).

\bibitem{SK36} B. Wunsch, T. Stauber, F. Sols and F. Guinea, Dynamical polarization of graphene at finite doping,  New J. Phys. {\bf 8}, 318 (2006).

\bibitem{Asgari} A. Qaiumzadeh and R. Asgari, Ground-state properties of gapped graphene using the random phase approximation, Phys. Rev. B {\bf 79}, 075414 (2009).

\bibitem{SK45} C. J. Tabert and E. J. Nicol, Dynamical polarization function, plasmons, and screening in silicene and other buckled honeycomb lattices, Phys. Rev. B {\bf 89}, 195410 (2014).

\bibitem{SK47} E. H. Hwang and S. Das Sarma, Dielectric function, screening, and plasmons in two-dimensional graphene,  Phys. Rev. B {\bf 75}, 205418 (2007).


\end{thebibliography}
\end{document}